\documentclass[%
reprint,
 aip,
 jap,
 longbibliography,
]{revtex4-1}

\usepackage{amsmath,amssymb,amsfonts}
\usepackage{graphicx}
\PassOptionsToPackage{hyphens}{url}
\usepackage{hyperref}
\usepackage{algorithm,algcompatible}
\usepackage{microtype}
\usepackage{dcolumn}
\usepackage{bm}
\usepackage{braket}
\usepackage{booktabs}

\usepackage[version=4]{mhchem}
\usepackage{xcolor}

\usepackage{silence}
\usepackage{subcaption}

\newcommand{\Var}[1]{\operatorname{Var}\left[#1\right]}
\newcommand{\Cov}[1]{\operatorname{Cov}\left[#1\right]}
\newcommand{\half}{\frac{1}{2}}
\newcommand{\sqrthalf}{\frac{1}{\sqrt{2}}}
\newcommand{\X}{\sigma^x}
\newcommand{\Y}{\sigma^y}
\newcommand{\Z}{\sigma^z}
\newcommand{\I}{I}
\newcommand{\Ketbra}[1]{\Ket{#1}\Bra{#1}}

\usepackage{amsthm}
\newtheorem{thm}{Theorem}

\graphicspath{{./fig/}}

\input{glyphtounicode}
\begin{document}

\title{Efficient evaluation of quantum observables using entangled measurements}

\author{Ikko Hamamura}
\email{hamamura@nucleng.kyoto-u.ac.jp}
\affiliation{%
  Kyoto University \\
  Department of Nuclear Engineering, Kyoto University, 6158540 Kyoto, Japan
}
\author{Takashi Imamichi}%
 \email{imamichi@jp.ibm.com}
\affiliation{%
  IBM Research -- Tokyo\\
  19-21, Nihonbashi Hakozaki-cho, Chuo-ku, 103-8510 Tokyo, Japan.
}%

\date{\today}

\begin{abstract}
  The advent of cloud quantum computing has led to the rapid development of quantum algorithms.
  In particular, it is necessary to study variational quantum-classical hybrid algorithms, which are executable on noisy intermediate-scale quantum (NISQ) computers.
  Evaluations of observables appear frequently in the variational quantum-classical hybrid algorithms for NISQ computers.
  By speeding up the evaluation of observables, it is possible to realize a faster algorithm and save resources of quantum computers.
  Grouping of observables with separable measurements has been conventionally used, and
  the grouping with entangled measurements has also been proposed recently by several teams.
  In this paper, we show that entangled measurements enhance the efficiency of evaluation of observables, both theoretically and experimentally
  by taking into account the covariance effect, which may affect the quality of evaluation of observables.
  We also propose using a part of entangled measurements for grouping to keep the depth of extra gates constant.
  Our proposed method is expected to be used in conjunction with other related studies.
  We hope that entangled measurements would become crucial resources, not only for joint measurements but also for quantum information processing.
\end{abstract}


\maketitle

\section*{Introduction}

It has been reported that many researchers have been working tirelessly to build a fault-tolerant quantum computer and quantum algorithms for years.
Recently, the development of quantum computing has been huge research attention.
The main reason is the rise of the noisy intermediate-scale quantum (NISQ) computers~\cite{Preskill2018},
which have 50--100 qubits, and its quantum error corrections are not yet implemented.
They are developed using superconducting~\cite{Kandala2017}
and trapped ion~\cite{Hempel2018, Lu2019} systems.
Although the number of qubits is small and the fidelity of operations is not very high,
programmable quantum computers have been made available, not only for researchers but also for public users.
For instance, IBM released IBM Q Experience in 2016, Rigetti released Quantum Cloud Services in 2018, and IonQ is going to start Quantum Cloud Service in 2019.
Software stacks for NISQ computers have also been extensively developed, for instance, Qiskit~\cite{Qiskit} by IBM, Forest~\cite{Smith2016} by Rigetti, and Cirq~\cite{Cirq} by Google.
Researchers are looking for killer applications for NISQ.
Quantum chemistry is one of the biggest targets~\cite{Peruzzo2014, Yung2014, Kandala2017, Grimsley2019}.
Optimization problem~\cite{Farhi2014, Guerreschi2019, Shaydulin2019} and
machine learning~\cite{Havlicek2019, Mitarai2018, Schuld2019, McClean2018} are also attractive applications.
For finance, some algorithms have also been proposed~\cite{Woerner2019, Stamatopoulos2019, Egger2019}.

In this paper, we focus on variational quantum eigensolver (VQE),
which is a quantum-classical hybrid algorithm proposed by Peruzzo et~al.~\cite{Peruzzo2014}
to compute eigenvalues and eigenvectors of matrices such as Hamiltonians.
VQE has been applied in various fields such as quantum chemistry~\cite{Moll2018}
and is extensively being studied
because NISQ computers can handle only short-depth circuits
and it is necessary to combine them with classical computers.
Such hybrid algorithms are relatively robust to noise compared with full-quantum algorithms.
Interested readers should refer to a review by McArdle~et~al.~\cite{McArdle2018} for details of VQE.

VQE minimizes the expectation value of the input operator by varying the quantum state $\ket{\psi(\theta)}$ with parameters $\theta$.
The expectation value of an operator $A$ with parameters $\theta$ can be expressed as $\Braket{A_\theta} = \braket{\psi(\theta)|A\psi(\theta)}$.
In quantum chemistry, an operator $A$ is usually a qubit Hamiltonian mapped from a fermionic Hamiltonian of molecules.
A qubit Hamiltonian can be written as a linear combination of tensor products of Pauli operators including the identity operator;
i.e., $A = \sum_{i=1}^n a_i P_i$,
where the tensor product of Pauli operators $P_i\in\Set{\X,\Y,\Z,\I}^{\otimes N}$ is referred to as \emph{Pauli string}.
VQE minimizes the expectation value by applying optimization algorithms for classical computers as follows:
\begin{align}
  \min_\theta \Braket{A_\theta} &= \min_\theta \braket{\psi(\theta)|A\psi(\theta)}\\
  &= \min_\theta \sum_{i=1}^{n} a_i \braket{\psi(\theta)|P_i\psi(\theta)}.
\end{align}
Quantum computers can be used to evaluate the expectation values of Pauli strings $\Braket{\psi(\theta)|P_i \psi(\theta)}$.

The quantum-classical hybrid algorithms require a large number of executions of quantum circuits to evaluate the expectation values of observables.
According to Wecker~et~al.~\cite{Wecker2015}, ``the required number of measurements is astronomically large for quantum chemistry applications to molecules.''
VQE consists of three nested iterations:
\begin{itemize}
	\item Outer iteration: to update the parameters $\theta$ of quantum state $\Ket{\psi(\theta)}$,
	\item Middle iteration: to evaluate the expectation value by calculating the weighted sum of Pauli strings, and
	\item Inner iteration: to evaluate the expectation value of a Pauli string through sampling.
\end{itemize}
The inner iteration evaluates Pauli strings as an expectation value with multiple samples.
This requires $O(\epsilon^{-2})$ samples for the statistical error $\epsilon$.
To reduce the inner iteration, Wang~et~al.~\cite{Wang2019} proposed another theoretical approach.

Herein, we focus on how to reduce the number of measurements in the middle iteration.
If Pauli strings are commutative, it implies they are compatible; i.e., they are jointly measurable.
McClean~et~al.~\cite{McClean2016} suggested a grouping of jointly measurable Pauli strings by using sequential measurements and pointed out the covariance effect.
Bravyi~et~al.~\cite{Bravyi2017} introduced the notion of grouping based on a tensor product basis (TPB).
Kandala~et~al.~\cite{Kandala2017} addressed the grouping of qubit Hamiltonians using TPB and analyzed the distribution of the standard error of its expectation value numerically.
Incompatibility by TPB can be represented by a graph called \emph{Pauli graph}.
It has been known that the grouping of Pauli strings can be reduced to the coloring problem of the Pauli graph.
Although the graph coloring problem is an NP-complete problem,
we can apply heuristic algorithms to obtain groups of Pauli strings,
e.g., the largest degree first coloring (LDFC) algorithm.

In this paper, we propose the use of entangled measurements in addition to TPB for the grouping of Pauli strings.
Entangled measurements are measurements of entangled observables.
Entangled observables are described by positive operator-valued measures that are not separable positive operator-valued measures~\cite{Hamamura2018}.
The advantage of using entangled measurements is that it makes it possible to achieve a smaller number of groups;
however, it is necessary to add extra CNOT gates to construct a measurement circuit corresponding to the group,
which may affect the fidelity of the resulting expectation value.
Therefore, we evaluate the properties of our grouping from both theoretical and experimental viewpoints.
Furthermore, we also propose a sampling strategy to mitigate the covariance effect by a group of qubit Hamiltonians.

The contributions of this paper are as follows:
\begin{enumerate}
  \item Grouping of Pauli strings with a part of entangled measurements,
  \item Measurement strategy based on the sizes of groups to suppress the covariance effects,
  \item Evaluation of the effect of the errors caused by additional CNOT gates, and
  \item Proof-of-concept demonstration of a simple Hamiltonian on real quantum computers.
\end{enumerate}

\section*{Results}
\subsection*{Evaluation of Pauli strings}

Let us consider the evaluation of the expectation values of quantum observables.
Target observables are mainly Hamiltonians.
Such observables of the multipartite qubit systems can be written as a linear combination of the Pauli strings
\(
  A = \sum_{i=1}^{n} a_i P_i,
\)
where $a_i$ denotes a real number and $P_i$ denotes an $N$-qubit Pauli string.
When the observables are not for the qubit systems,
the Jordan--Wigner transformation for fermions or the Jordan--Schwinger transformation for bosons can be used to map the qubit systems from other systems.
Herein, we present the transformations for fermions in more detail.
The second quantized fermionic Hamiltonian can be expressed as
\begin{equation}
  H = \sum_{ij} h_{ij} c_i^\dag c_j + \frac{1}{2}\sum_{ijkl} h_{ijkl} c_i^\dag c_j^\dag c_l c_k,
\end{equation}
where $h_{ij}$ denotes kinetic and potential energy and $h_{ijkl}$ denotes interaction.
Herein, $c_i$ denotes an annihilation operator of the $i$-th fermion, and $c_j^\dag$ denotes a creation operator of the $j$-th fermion.
There are three well-known transformations from fermionic systems to qubit systems:
the Jordan--Wigner transformation~\cite{Jordan1928}, that was originally proposed for lattice systems, parity transformation, and Bravyi--Kitaev transformation~\cite{Bravyi2002, Seeley2012}.
Note that the difference between transformations can affect the results of grouping.

When one prepares a quantum state $\ket{\psi}$ in a quantum computer, the expectation value of a quantum observable $A$ can be expressed as $\braket{A} = \braket{\psi|A\psi}$.
The expectation values of Pauli operators can be evaluated as follows.
First, a measurement in the computational basis is available in a quantum computer.
Then, by employing a single-qubit unitary rotation before the measurements, we can implement the measurements of other bases.
Calculation
\(
  \Braket{A} = \sum_{i=1}^{n} a_i \Braket{P_i}
\)
gives the expectation value of $A$.

\subsection*{Grouping Pauli strings with TPB}

Some Pauli strings are simultaneously diagonalizable by using the tensor product of bases $\set{\mathcal{X},\mathcal{Y}, \mathcal{Z}}$
where $\mathcal{X} = \left\lbrace\ket{0}\pm\ket{1}\right\rbrace$,
$\mathcal{Y}=\left\lbrace\ket{0}\pm i\ket{1}\right\rbrace$,
and $\mathcal{Z} = \left\lbrace\ket{0},\ket{1}\right\rbrace$.
Herein, the sets of bases are referred to as TPB sets.
This implies that some Pauli strings that are simultaneously diagonalizable by using TPB are jointly measurable by TPB.
This also implies that such Pauli strings can be grouped and we can obtain their expectation values at the same time.

We introduce the Pauli graph $G = (V, E)$ of an observable $A=\sum_i a_i P_i$ as follows:
nodes $V$ correspond to Pauli strings $P_i$ and edge $(u, v) \in E$ spans if Pauli strings $u$ and $v$ are \emph{not} jointly measurable by TPB.
Further, a coloring of the Pauli graph gives groups of the Pauli strings that are jointly measurable.
For instance, Qiskit~\cite{Qiskit} adopts the LDFC algorithm,
which first sorts the nodes in descending order of degree and then assigns the smallest color number not used by its colored neighbors.
The details of the Pauli graph and LDFC algorithm are presented in Appendix~A of the Supplementary Information.
It executes fast for practical graphs of interest,
and the number of the resulting groups is close to the lower bound by the max clique of the Pauli graph
(see Table~\ref{table:num-groups} for details).

\subsection*{Grouping with TPB and entangled measurements}

Measurements by TPB are separable measurements.
We propose taking advantage of entangled measurements.
We introduce a new grouping approach for Pauli strings
that uses not only TPB but also entangled measurements such as Bell measurements
to reduce the number of measurements.

For instance, the expectation values of $\X\X$, $\Y\Y$, and $\Z\Z$ cannot be obtained simultaneously from TPB measurements.
It requires three types of measurements to compute the expectation values of $\X\X$, $\Y\Y$, and $\Z\Z$;
however, these Pauli strings can be measured jointly using Bell measurement (see the Methods section for details).

Simultaneous diagonalization provides a joint measurement using entangled observables.
The incompatibility of Pauli strings can be checked by parity of the number of different Pauli operators $\X, \Y$, and $\Z$.
It is possible to construct an \emph{extended Pauli graph} based on this incompatibility and calculate the number of groups using the LDFC algorithm,
which we refer to as ``ALL'' in the subsequent sections.
We will calculate the number of groups using all measurements and LDFC later (Table~\ref{table:num-groups}).
However, the computational cost of the circuit construction and the circuit depth increase according to the size of entanglements.
Therefore, it may not be practical for NISQ computers to use all measurements
due to the fidelity of operations of NISQ computers, especially multi-qubit operations.

To mitigate the drawback, we propose another approach that involves the use of a part of the entanglement of the observables.
It consists of two phases: choosing a set of entangled observables and grouping of Pauli strings with TPB and the set of entangled observables.

The first phase is to choose entangled measurements (e.g., Bell measurements and omega measurements~\cite{Entanglion}).
We would present details of the Bell measurements in the Method section and present other two-qubit entangled measurements in Appendix~B of the Supplementary Information.
One constructs quantum circuits corresponding to the entangled measurements.
This task can be done by using simultaneous diagonalization as a preprocessing technique.
We can alternatively use other methods based on Clifford gates, which was proposed recently~\cite{Jena2019, Yen2019, Gokhale2019}.
Because quantum circuits can be generated in advance in this phase, therefore, the cost of circuit construction in the next phase can be reduced.

The second phase is to construct groups of Pauli strings using TPB and the entangled measurements chosen in the first phase.
It is possible to construct a Pauli graph in the same way as using TPB-based methods and sort the nodes in descending order of degree.
Then, the nodes can be merged if they are jointly measurable by using TPB and the entangled measurements.
It is noteworthy that this joint measurability depends on merge history.
If one merges nodes, it is necessary to use the entangled measurement at particular positions of Pauli strings,
and after that, it is not possible to change the measurement.
The merged nodes correspond to groups of measurements, and
one can construct quantum circuits corresponding to the groups and perform the measurements.
We refer to the grouping that involves the use of TPB and Bell measurements as ``TPB+Bell,''
and the grouping with TPB and all two-qubit measurements as ``TPB+2Q.''
Refer to the Methods section for details of the grouping algorithm.

Entanglement of observables could be considered as a resource of a joint measurement.
In experiments with NISQ computers, available entangled measurements are limited because of multi-qubit gate errors that may occur.
Therefore, we need to consider algorithms with available measurements.
For instance, if one uses two-qubit entangled measurements, only one entangler (e.g., CNOT gate) is required for each of the qubits.
Our proposed method requires constant depth measurement circuits for fixed entangled measurements.
The advantage of our approach is that we can adjust the depth of the additional circuits for entangled measurements
by considering errors due to multi-qubit gates.

\subsection*{Standard error caused by the grouping of Pauli strings}
McClean et~al.~\cite{McClean2016} discussed covariance effects and showed that the additional covariance caused as a result of grouping may require more measurements.
However, Kandala et~al.~\cite{Kandala2017} numerically verified that the grouping with TPB has fewer errors than the no-grouping strategy for some molecules.
It is noteworthy that the square of the error is in inverse proportion to the number of samples.
We show mathematically that if the number of samples in each group is in proportion to the size of the group,
the standard error of a grouping is smaller than that of no-grouping in Appendix~C of the Supplementary Information, where the total numbers of samples for the grouping and no-grouping strategies are the same.

\subsection*{Number of groups}

\begin{table*}
  \centering
  \caption{%
  Comparison of the numbers of groups
}
  \label{table:num-groups}
  \begin{tabular}{ll|rrrrr|r}
    Molecule & Trans-& \multicolumn{5}{c|}{Number of groups}&Clique size of\\
    & formation& No-grouping & TPB & TPB+Bell & TPB+2Q & ALL & Pauli graph\\
\midrule
&JW&631&136&42&42&35			&130$^\ast$\\
\ce{LiH}&Parity&631&165&72&89&35		&160$^\ast$\\
&BK&631&211&103&113&35			&208$^\ast$\\
\midrule
&JW&1150&215&59&59&58			&200$^\ast$\\
\ce{BeH2}&Parity&1150&323&106&124&58	&313$^\ast$\\
&BK&1150&341&199&200&58		&335$^\ast$\\
\midrule
&JW&1858&380&71&71&84			&355$^\ast$\\
\ce{H2O}&Parity&1858&495&130&199&82		&482$^\ast$\\
&BK&1858&515&286&273&82			&508$^\ast$\\
\midrule
&JW&4973&1052&127&127&117		&994$^\ast$\\
\ce{NH3}&Parity&4973&1091&205&360&115	&955$\phantom{^\ast}$\\
&BK&4973&1086&579&538&115		&1040$\phantom{^\ast}$\\
\midrule
&JW&4427&906&154&153&110		&844$^\ast$\\
\ce{HCl}&Parity&4427&1098&294&418&112	&1047$^\ast$\\
&BK&4427&1434&733&686&112		&1413$^\ast$\\
  \end{tabular}
\end{table*}

We applied four types of grouping methods
(one with TPB; one with TPB and Bell measurements; one with TPB and all two-qubit entangled measurements; and one with all measurements)
to the following molecules: \ce{LiH}, \ce{BeH2}, \ce{H2O}, \ce{NH3}, and \ce{HCl}.
We also computed the maximal clique sizes of Pauli graphs for TPB by applying the MCQD algorithm~\cite{Konc2007}.
We ran MCQD on Intel Xeon E5-2690 CPU with a 1-hour time limit
and were able to observe the maximum cliques for all Pauli graphs except \ce{NH3} Parity and Bravyi--Kitaev.
Note that the Hamiltonians of the molecules are presented as ancillary files in a previous study~\cite{Bravyi2017}.

We present a comparison of the numbers of groups in Table~\ref{table:num-groups}.
  The clique sizes with `$\ast$' are the maximum.
  JW and BK are Jordan--Wigner and Bravyi--Kitaev transformations, respectively.
  TPB, TPB+BELL, TPB+2Q, and ALL denote the groupings using TPB, TPB and Bell measurements,
  TPB and all two-qubit entangled measurements, and all measurements, respectively.
The grouping with TPB yielded several groups that are slightly larger than the clique sizes of the Pauli graphs.
It should be noted that the grouping with TPB cannot yield fewer numbers of groups than the clique size
because it is based on a coloring algorithm (see Eq.~64.1~\cite{Schrijver2003}).
On the other hand, the groupings using entangled measurements (TPB+Bell, TPB+2Q, and ALL) achieved fewer numbers than the clique sizes.
The grouping with TPB and two-qubit entangled measurements (TPB+2Q) does not always result in a fewer number of groups than those obtained by grouping with TPB and Bell measurements (TPB+Bell).
This implies that our heuristic algorithm cannot take full advantage of more choices of measurements.
We also observed that the grouping with all measurements (ALL) does not always result in the smallest number of groups among all methods.
This implies that LDFC cannot be used to obtain good solutions for the extended Pauli graphs.
The number of groups depends on the type of transformations as well as the grouping methods used.
For instance, for Jordan--Wigner transformation, TPB+Bell results in a smaller number of groups for most molecules than groupings using TPB and TPB+2Q.

We observed that the entangled measurements are effective in reducing the number of measurements of Pauli strings.
However, there exist various errors in using NISQ computers, especially the two-qubit gate error that is generally much larger than a one-qubit error.
In the next section, we discuss the effects of additional CNOT gates introduced in entangled measurements.

\subsection*{Effect of additional CNOT gates}

Entangled measurements require additional two-qubit gates.
We evaluate the effect of additional CNOT gates for a Bell measurement in one of the simplest models, two-qubit antiferromagnetic Heisenberg model,
whose Hamiltonian can be expressed as
\begin{equation}
  H = \X\X + \Y\Y + \Z\Z. \label{eq:afm}
\end{equation}
Herein, we set the coupling constant to be $1$ for simplicity.
These Pauli strings cannot be grouped by using TPB but can be grouped by using a Bell measurement.
We compute the expectation value with the ground state using a noise model in Qiskit Aer and readout error mitigation~\cite{Temme2017} in Qiskit Ignis.
We assigned 2000 samples to each group of no-grouping (three groups) and 6000 samples to a group of the grouping obtained by using a Bell measurement (one group).
We assume that the one-qubit readout errors are $p(1|0) = 0.01$ and $p(0|1) = 0.1$, and the one-qubit depolarizing error is 0.001.
We simulated the expectation value of the Hamiltonian by varying the two-qubit depolarizing error.
See Table~\ref{tb:cnoterror} for the results.
The row ``No-grouping (raw)'' denotes the expectation values with the standard error,
where we did not apply the readout error mitigation.
On the other hand, ``No-grouping (mit)'' denotes the results obtained when the readout error mitigation was applied.
The rows of ``Bell (raw)'' and ``Bell (mit)'' show the results of the grouping of a Bell measurement with and without the readout error mitigation, respectively.
We observed that the grouping with a Bell measurement achieved comparable expectation value with less standard error
compared with no-grouping up to two-qubit error 0.010.
The grouping with a Bell measurement resulted in worse expectation values with two-qubit error more than 0.010
owing to the extra CNOT gate introduced by using the Bell measurement.
It should be noted that we implemented our program with Qiskit Terra 0.7.2, Qiskit Aer 0.1.1, Qiskit Ignis 0.1.0, and Qiskit Aqua 0.4.1.
\begin{table*}[htbp]
  \centering
  \caption{%
	Simulation results of the expectation values and standard errors of the two-qubit antiferromagnetic Heisenberg model
	}\label{tb:cnoterror}
  \begin{tabular}{l|ccccc}
	Two-qubit error&0&0.005&0.010&0.015&0.020\\
	\midrule
	No-grouping (raw)&$-2.318\pm0.043$&$-2.327\pm0.042$&$-2.313\pm0.043$&$-2.311\pm0.043$&$-2.293\pm0.043$\\
	No-grouping (mit)&$-2.970\pm0.008$&$-2.971\pm0.007$&$-2.959\pm0.009$&$-2.957\pm0.009$&$-2.944\pm0.011$\\
	Bell (raw)&$-2.227\pm0.020$&$-2.233\pm0.020$&$-2.207\pm0.021$&$-2.186\pm0.021$&$-2.153\pm0.021$\\
	Bell (mit)&$-2.966\pm0.005$&$-2.987\pm0.003$&$-2.956\pm0.005$&$-2.929\pm0.007$&$-2.889\pm0.008$\\
  \end{tabular}
\end{table*}

We also evaluated additional CNOT gate effects using real quantum computers, IBM Q Tokyo and IBM Q Poughkeepsie (Table~\ref{tb:cnotexp}).
Note that both have 20 qubits.
We picked out the pair of qubits with the best fidelity based on the property data of the devices
when we executed the circuits.
The properties of the devices, including connectivity and error distributions, can be found in the literature\cite{Corcoles2019}.
We observed that the grouping obtained using the Bell measurement achieved comparable expectation value with less standard error
compared with the no-grouping type.
These experiments suggest that readout error has a significant influence (more than the additional two-qubit error).
\begin{table}[htbp]
  \centering
  \caption{Experiment on the additional qubit gates using IBM Q Tokyo and IBM Q Poughkeepsie}\label{tb:cnotexp}
  \begin{tabular}{l|cc}
    Device name & Tokyo & Poughkeepsie \\
    \midrule
    No-grouping (raw) &  $-2.588\pm0.034$ & $-2.313 \pm 0.043$ \\
    No-grouping (mit) & $-2.962 \pm0.010$ & $-2.889 \pm0.017$ \\
    Bell (raw) & $-2.487 \pm0.017$ & $-2.432 \pm0.018$ \\
    Bell (mit) & $-2.984 \pm0.003$ & $-2.983 \pm0.003$
  \end{tabular}
\end{table}

\subsection*{VQE with entangled measurements}

\begin{figure*}[ht]
  \centering
  \includegraphics[width=.9\linewidth]{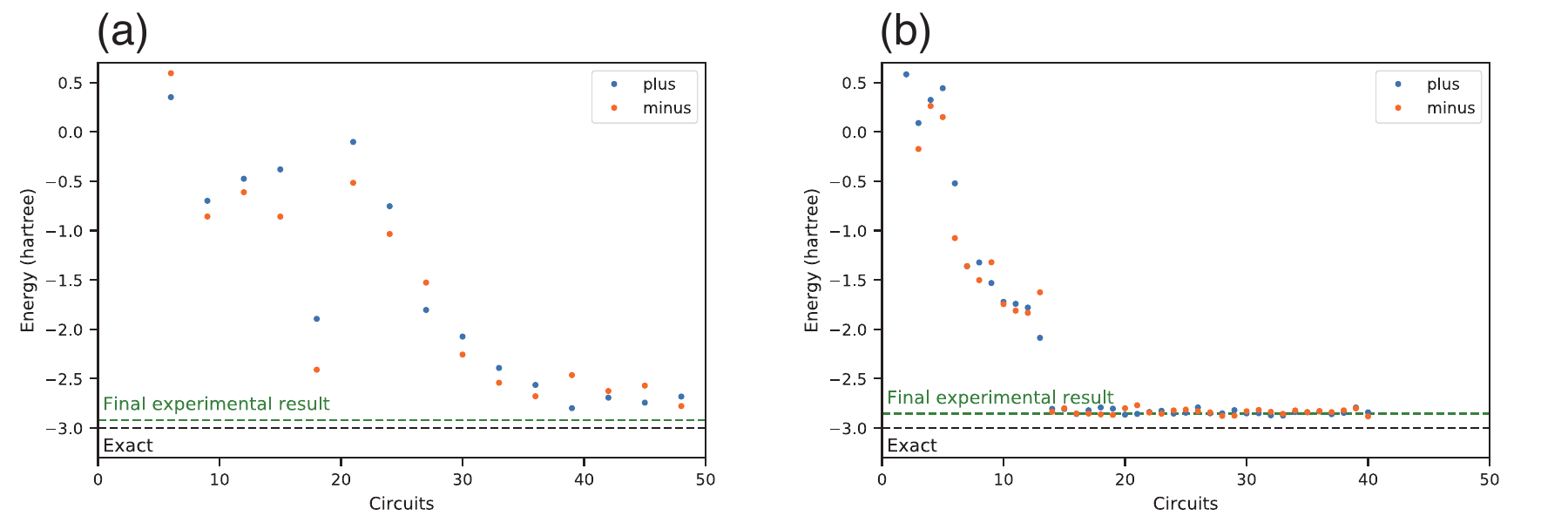}
  \caption{%
    Comparison of VQE with no-grouping (a) and one with the grouping obtained using a Bell measurement (b).
    The horizontal and the vertical axes represent the number of quantum circuits used and expectation values of the Heisenberg Hamiltonian, respectively.
    The graph legends (plus and minus) are used to minimize the expectation value in the SPSA algorithm.
  	The black dashed line and the green one denote the exact ground state energy and the final result of the VQE experiment, respectively.
  }
  \label{fig:vqe-heisenberg}
\end{figure*}

We incorporate the grouping with Bell measurements into VQE
to calculate the ground state energy of the two-qubit antiferromagnetic Heisenberg model~\eqref{eq:afm},
and we use the no-grouping as the baseline.
We compared the number of circuits to be converged.
We used $Ry$ trial wave function with depth 1 as the variational form and
the simultaneous perturbation stochastic approximation (SPSA)~\cite{Spall1992} as the optimizer.
We executed VQE algorithm with 40 iteration steps.
We applied the readout error mitigation and calibrated the mitigation data every 10 iteration steps of VQE.
We fixed the number of executions of quantum circuits to 8192 samples, i.e., the maximum number for a job of the IBM Q systems as of July 2019.
See Fig.~\ref{fig:vqe-heisenberg} for the results.
The horizontal axis and the vertical axis represent the number of quantum circuits used and expectation values of the Hamiltonian, respectively.
VQE with a Bell measurement used one circuit for each iteration, whereas VQE without grouping used three circuits for each iteration.
We plotted the standard error as well as the expectation values,
but the error bar is small enough to be overshadowed by point.
We observed that VQE with a Bell measurement required fewer circuits than VQE with no-grouping to converge.
We will present more results and discuss them in Appendix~D of the Supplementary Information.

\section*{Discussion}

In this paper, we presented the efficient evaluation of Pauli strings with a part of entangled measurements.
We propose reducing the number of groups of Pauli strings that are jointly measurable by using both TPB and a part of entangled measurements.
It is remarkable that grouping by using TPB and Bell measurements (TPB+Bell) produces as few number of groups as
the grouping obtained by using all measurements (ALL) for some molecules with the Jordan--Wigner transformation as shown in Table~\ref{table:num-groups}
as the depth of the additional CNOT gates required for TPB+Bell is constant, 1.
Jordan--Wigner transformation has an even number of Pauli $\X$ and $\Y$ for the fermion Hamiltonian.
This even structure does not appear in Bravyi--Kitaev and parity transformations.
This property might be fit for the Bell measurement,
which is a joint measurement of $\X\X$, $\Y\Y$, and $\Z\Z$.
It should be noted that grouping using all measurements were obtained by applying LDFC to the extended Pauli graph;
thus, the solutions may be suboptimal, and they can be improved by using more sophisticated coloring algorithms.
We also showed that if the number of samples in each group is in proportion to the size of the group,
the standard error obtained with grouping is always equal to or smaller than that with no-grouping.

We discuss the connectivity of NISQ computers.
There are two types of NISQ computers.
One has all-to-all connectivity, such as a trapped ion system.
For full connectivity devices, there is no connectivity problem.
The other has nearest neighbor connectivity, such as superconducting systems.
There are two strategies to apply our proposed method:
(1) one limits the grouping algorithm to adapt the connectivity or
(2) one changes the initial layout of the qubits.
For the latter, our proposed method using two-qubit measurements needs one nonlocal gate per two qubits and then low connectivity.
For example, a one-dimensional lattice is sufficient to map the measurement circuits.

Public quantum computers are available as cloud service today.
The computation time is bounded by job time, and we refer to it as \emph{job bound}.
The time needed for a job to be completed is equal to the sum of waiting and execution time.
The execution time can be regarded as the execution time of the circuits included in the job.
In other words, execution time is roughly proportional to the total number of circuits to be executed.
Therefore, reducing the number of circuits in grouping contributes to the improvement of execution time.

In the present study, we implement the entangled measurements using the Heisenberg picture in the software layer.
There are other possible candidates for implementing joint measurements.
The entangled measurements can be implemented in the hardware layer~\cite{Chow2010}, called joint readout.
The joint readout may make our proposal more accurate and precise.
Direct or indirect sequential measurements also provide joint measurements.

To accelerate the iterations of VQE, several approaches have been discussed.
An efficient partitioning using mean-field approach was proposed by Izmaylov et~al.~\cite{Izmaylov2019}
It requires feedforward measurements, which are not implemented in the current devices.
However, once the feedforward is implemented, we may achieve better efficiency when it is used in combination with our proposed methods.
A sequential minimal optimization method~\cite{Nakanishi2019} reduces the outer iteration.
The proposed method converges faster than previous optimizers and uses no hyperparameter.
Fermionic representability conditions can be used to reduce the middle iterations~\cite{Rubin2018}.
Some efficiency improvements that do not relate to iterations were studied~\cite{Setia2018, Babbush2018, Barkoutsos2018, Romero2018, Gard2019}.
Our proposed method can be used together with them.

Let us briefly review the history of the grouping based on TPB.
Kandala~et~al.~\cite{Kandala2017} used the grouping by TPB in their experiments on some molecules.
Qiskit~\cite{Qiskit} presented the implementation of a coloring-based grouping algorithm of the Pauli graph in June 2018,
which was released as QISKit ACQUA 0.1.0.
The grouping by TPB was also implemented in the OpenFermion~\cite{OpenFermion} v0.9.0 (December 2018) and PyQuil~\cite{Smith2016} 2.2 (January 2019).
Verteletskyi et~al.\cite{Verteletskyi2019} systematically studied the heuristic algorithms for grouping using the TPB.

Recently, many studies to enhance the evaluation of observables have been presented~\cite{Jena2019, Yen2019, Izmaylov2019b, Huggins2019, Gokhale2019, Crawford2019,Zhao2019,Gokhale2019b}.
Entangled measurements have been used either implicitly or explicitly in some studies~\cite{Jena2019, Yen2019, Gokhale2019,Crawford2019,Zhao2019,Gokhale2019b}.
In contrast to these studies that use all measurements, we employ TPB and a part of entangled measurements.
Jena et~al.~\cite{Jena2019} showed that grouping of Pauli strings is NP-hard in general qudit system.
They discussed the case when the available gate set is limited to Clifford gates or single-qudit Clifford gates, whereas we discuss the case where available measurements are limited.
The diagonalization by Clifford operators corresponds to all measurements, and the diagonalization by using single-qubit Clifford corresponds to TPB.
Yen~et~al.~\cite{Yen2019} proposed a conversion method from groups of commuting Pauli strings into TPB using unitary transformations.
Their unitary transformations were discussed in the Schr\"{o}dinger picture, but they used all measurements in the Heisenberg picture
and proposed efficient circuit construction using Clifford gates.
Gokhale~et~al.\cite{Gokhale2019} used entangled measurements to measure commuting Pauli strings
and developed a circuit synthesis tool for joint measurement based on the stabilizer.
Furthermore, their results have been demonstrated experimentally in computing the ground state energy of deuteron on the IBM Q Tokyo.
Crawford~et~al.~\cite{Crawford2019} proposed taking into account coefficients of Pauli strings for the grouping to mitigate the covariance effect,
whereas we propose an adjustment of the number of samples while taking into account the sizes of groups.
A linear reduction of the number of groups from $O(N^4)$ to $O(N^3)$ for molecular Hamiltonians has been proven in different ways~\cite{Zhao2019,Gokhale2019b}.
Izmaylov~et~al.~\cite{Izmaylov2019b} and Huggins~et~al.~\cite{Huggins2019} developed different methods that are not based on a joint measurement of the Pauli strings.
  Jiang~et~al.~\cite{Jiang2019} used a Bell measurement on a system and an ancilla to implement a symmetric informationally complete POVM,
  whereas our proposal uses a Bell measurement on systems.

\section*{Methods}

\subsection*{Entangled measurements}

Herein, we introduce entangled measurements.
Entanglement is a specific property of quantum theory and an essential concept in quantum information science and technology.
Entanglement is originally defined for quantum states.
Bell states are one of the maximally entangled states.
Bell states can be defined as follows:
\begin{align}
  \Ket{\Phi^\pm} = \sqrthalf \Ket{00} \pm \sqrthalf \Ket{11},
  \Ket{\Psi^\pm} = \sqrthalf \Ket{01} \pm \sqrthalf \Ket{10}. \nonumber
\end{align}

\begin{figure}[ht]
  \centering
  \includegraphics[height=10em]{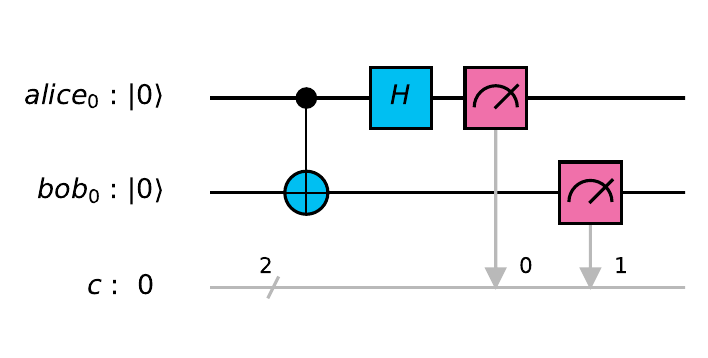}
  \caption{%
  Quantum circuit of the Bell measurement.
  This is an inverse process of creating Bell states.
}
  \label{fig:bell}
\end{figure}

Entanglement of states can be extended to observables.
As entanglement of states can be detected by the violation of Bell-CHSH inequality~\cite{Bell1964, Clauser1969},
an entanglement of observables can be detected by the violation of the dual Bell-CHSH inequality~\cite{Hamamura2018}.
Measurements of entangled observables are referred to as entangled measurements.
Let us introduce examples of entangled measurements.
The first example of entangled measurements is Bell measurements defined as the projective measurements on Bell states.
The Bell measurements are used in the quantum teleportation protocol~\cite{Bennett1993}.
A remarkable property of the Bell measurement is that expectation values of $\X\otimes\X$, $\Y\otimes\Y$, and $\Z\otimes\Z$ can be calculated from the result of the Bell measurement.
This property is based on the following fact:
\footnotesize
\begin{align*}
  \X\otimes\X &= \Ketbra{\Psi^+} + \Ketbra{\Phi^+} - \Ketbra{\Psi^-} - \Ketbra{\Phi^-}, \\
  \Y\otimes\Y &= \Ketbra{\Psi^+} + \Ketbra{\Phi^-} - \Ketbra{\Phi^+} - \Ketbra{\Psi^-}, \\
  \Z\otimes\Z &= \Ketbra{\Phi^+} + \Ketbra{\Phi^-} - \Ketbra{\Psi^+} - \Ketbra{\Psi^-}.
\end{align*}
\normalsize
CNOT and Hadamard gates are needed to implement the Bell measurement on a quantum computer.
The quantum circuit is shown in Fig.~\ref{fig:bell}.
We will describe other useful entangled measurements in Appendix~B of the Supplementary Information.

\subsection*{Grouping algorithm}
We propose a heuristic algorithm of grouping Pauli strings using TPB and a part of entangled measurements greedily.
We first make a Pauli graph as the grouping with TPB does and then merge nodes with high degrees if they are jointly measurable by given entangled measurements.
To check whether a pair of Pauli strings is compatible, we generate permutations of qubit positions and check if any of the entangled measurements can be applied to the position.
The number of resulting groups depends on the order of measurements to be applied in Algorithm~\ref{alg:assign_measure}.
We let $E = \{\text{Bell}, \text{TPB}\}$ for `TPB+Bell' and
$E = \{\text{Bell}, \text{Omega-XX}, \text{Omega-YY}, \text{Omega-ZZ}, \text{Chi}, \text{TPB}\}$ for `TPB+2Q' in our experiments.
See Algorithms~\ref{alg:group_ent} and~\ref{alg:assign_measure} for details of the algorithm
and Appendix~B of the Supplementary Information for the definitions of the omega and chi measurements.
\begin{algorithm}[H]
	\caption{Greedy grouping of Pauli strings with available measurements}
	\label{alg:group_ent}
	\begin{algorithmic}[1]
		\STATE \textbf{Input}: observable $A = \sum_{i=1}^n a_i P_i$, a set of measurements $E = \set{\mathcal{E}_1,\dots,\mathcal{E}_l}$. Note that each Pauli string has length $N$ and the measurements may include both TPB and entangled measurements.
		\STATE Make a Pauli graph $G$ of Pauli strings $\set{P_i}$ based on TPB.
		\STATE Sort the nodes in the descending order of degree on $G$ and let them be $\set{v_1,\dots,v_n}$
		\STATE Initialize the assignment of measurements $M_i$ of all nodes $v_i$.
		\FOR{$i = 1,\dots,n$}
			\STATE \textbf{if} $v_i$ is already merged, skip it.
			\FOR{$j = i + 1, \dots, n$}
				\STATE \textbf{if} $v_j$ is already merged, skip it.
				\IF{$v_i$ and $v_j$ are jointly measurable by $M_i$ and $E$ (see Algorithm~\ref{alg:assign_measure})}
					\STATE Merge $v_j$ to $v_i$ and update $M_i$.
				\ENDIF
			\ENDFOR
		\ENDFOR
		\STATE Return the merged nodes as the groups of observables.
	\end{algorithmic}
\end{algorithm}
\begin{algorithm}[H]
	\caption{Greedy assignment of measurements}
	\label{alg:assign_measure}
	\begin{algorithmic}[1]
		\STATE \textbf{Input}: a pair of Pauli strings $(v_i, v_j)$, a set of measurements $E$, current assignment of measurements $M_i$ for $v_i$.
		\STATE Note that $M_i$ represents a set of measurements and the qubits associated with the measurements, e.g., first qubit to be measured by $\X$ and second and fourth qubits to be measured by $\Y\Y$.
		\STATE Return \textbf{fail} if any qubit of $v_j$ is not compatible with $M_i$.
		\STATE Let $U$ be the set of positions of qubits that $M_i$ does not cover.
		\STATE Remove positions of qubits from $U$ where $v_i$ and $v_j$ have common Pauli operators.
		\WHILE{$U \neq \emptyset$} \label{for:top}
			\FOR{$\mathcal{E} \in E$}
		\FOR{$p \in$ permutation of $U$ with the length of measurement $\mathcal{E}$}
		\IF{both $v_i$ and $v_j$ are compatible with $\mathcal{E}$ at position $p$}
						\STATE Update $M_i$ with $\mathcal{E}$ at position $p$.
						\STATE Update $U$ by removing $p$.
						\STATE Go to line \ref{for:top}.
					\ENDIF
				\ENDFOR
			\ENDFOR
			\STATE Return \textbf{fail}.
		\ENDWHILE
		\STATE Return $M_i$.
	\end{algorithmic}
\end{algorithm}

\section*{Data Availability}
The data of the plots of VQE that support the finding of this study are available from the corresponding authors upon reasonable request.

\section*{Code Availability}
The software implementation used to group the Pauli strings and perform entangled measurements is not publicly available.

\section*{ACKNOWLEDGMENTS}

We thank Sergey Bravyi, Antonio Mezzacapo, Rudy Raymond, and Toshinari Itoko for fruitful discussions
and Ryo Takakura, Naixu Guo, and Kazuki Yamaga for helpful comments on the manuscript.
I.\ H.\ has done part of this study during the internship at IBM Research -- Tokyo in 2018.
I.\ H.\ also acknowledges support by Grant-in-Aid for JSPS Research Fellow (JP18J10310).

\section*{COMPETING INTERESTS}
A part of this work was included in a patent filed with the Japan Patent Office and that is going to be filed with the US Patent and Trademark office.

\section*{AUTHOR CONTRIBUTIONS}
I.\ H.\ and T.\ I.\ performed the theoretical analysis and experiments.
I.\ H.\ proved the theorem about the standard error.
I.\ H.\ and T.\ I.\ contributed to write the paper.

\bibliographystyle{naturemag}
\bibliography{ref}

\begin{thebibliography}{10}
\expandafter\ifx\csname url\endcsname\relax
  \def\url#1{\texttt{#1}}\fi
\expandafter\ifx\csname urlprefix\endcsname\relax\def\urlprefix{URL }\fi
\providecommand{\bibinfo}[2]{#2}
\providecommand{\eprint}[2][]{\url{#2}}

\bibitem{Preskill2018}
\bibinfo{author}{Preskill, J.}
\newblock \bibinfo{title}{Quantum {C}omputing in the {NISQ} era and beyond}.
\newblock \emph{\bibinfo{journal}{{Quantum}}} \textbf{\bibinfo{volume}{2}},
  \bibinfo{pages}{79} (\bibinfo{year}{2018}).
\newblock \urlprefix\url{https://doi.org/10.22331/q-2018-08-06-79}.

\bibitem{Kandala2017}
\bibinfo{author}{Kandala, A.} \emph{et~al.}
\newblock \bibinfo{title}{Hardware-efficient variational quantum eigensolver
  for small molecules and quantum magnets}.
\newblock \emph{\bibinfo{journal}{Nature}} \textbf{\bibinfo{volume}{549}},
  \bibinfo{pages}{242--246} (\bibinfo{year}{2017}).
\newblock \urlprefix\url{http://dx.doi.org/10.1038/nature23879}.

\bibitem{Hempel2018}
\bibinfo{author}{Hempel, C.} \emph{et~al.}
\newblock \bibinfo{title}{{Quantum Chemistry Calculations on a Trapped-Ion
  Quantum Simulator}}.
\newblock \emph{\bibinfo{journal}{Phys. Rev. X}} \textbf{\bibinfo{volume}{8}},
  \bibinfo{pages}{031022} (\bibinfo{year}{2018}).
\newblock \urlprefix\url{https://link.aps.org/doi/10.1103/PhysRevX.8.031022}.

\bibitem{Lu2019}
\bibinfo{author}{Lu, Y.} \emph{et~al.}
\newblock \bibinfo{title}{Global entangling gates on arbitrary ion qubits}.
\newblock \emph{\bibinfo{journal}{Nature}}  (\bibinfo{year}{2019}).
\newblock \urlprefix\url{https://doi.org/10.1038/s41586-019-1428-4}.

\bibitem{Qiskit}
\bibinfo{title}{{Qiskit: An Open-source Framework for Quantum Computing}}.
\newblock
  \bibinfo{howpublished}{\href{https://doi.org/10.5281/zenodo.2562110}{doi:10.5281/zenodo.2562110}}
  (\bibinfo{year}{2019}).

\bibitem{Smith2016}
\bibinfo{author}{Smith, R.~S.}, \bibinfo{author}{Curtis, M.~J.} \&
  \bibinfo{author}{Zeng, W.~J.}
\newblock \bibinfo{title}{{A Practical Quantum Instruction Set Architecture}}
  (\bibinfo{year}{2016}).
\newblock \eprint{arXiv:1608.03355}.

\bibitem{Cirq}
\bibinfo{title}{{Cirq: A python framework for creating, editing, and invoking
  Noisy Intermediate Scale Quantum (NISQ) circuits}}.
\newblock \bibinfo{howpublished}{\url{https://github.com/quantumlib/Cirq}}
  (\bibinfo{year}{2018}).

\bibitem{Peruzzo2014}
\bibinfo{author}{Peruzzo, A.} \emph{et~al.}
\newblock \bibinfo{title}{A variational eigenvalue solver on a photonic quantum
  processor}.
\newblock \emph{\bibinfo{journal}{Nat. Commun.}} \textbf{\bibinfo{volume}{5}},
  \bibinfo{pages}{4213} (\bibinfo{year}{2014}).
\newblock \urlprefix\url{http://dx.doi.org/10.1038/ncomms5213}.

\bibitem{Yung2014}
\bibinfo{author}{Yung, M.~H.} \emph{et~al.}
\newblock \bibinfo{title}{From transistor to trapped-ion computers for quantum
  chemistry}.
\newblock \emph{\bibinfo{journal}{Sci. Rep.}} \textbf{\bibinfo{volume}{4}},
  \bibinfo{pages}{3589} (\bibinfo{year}{2014}).
\newblock \urlprefix\url{https://doi.org/10.1038/srep03589}.

\bibitem{Grimsley2019}
\bibinfo{author}{Grimsley, H.~R.}, \bibinfo{author}{Economou, S.~E.},
  \bibinfo{author}{Barnes, E.} \& \bibinfo{author}{Mayhall, N.~J.}
\newblock \bibinfo{title}{An adaptive variational algorithm for exact molecular
  simulations on a quantum computer}.
\newblock \emph{\bibinfo{journal}{Nat. Commun.}} \textbf{\bibinfo{volume}{10}},
  \bibinfo{pages}{3007} (\bibinfo{year}{2019}).
\newblock \urlprefix\url{https://doi.org/10.1038/s41467-019-10988-2}.

\bibitem{Farhi2014}
\bibinfo{author}{Farhi, E.}, \bibinfo{author}{Goldstone, J.} \&
  \bibinfo{author}{Gutmann, S.}
\newblock \bibinfo{title}{{A Quantum Approximate Optimization Algorithm}}
  (\bibinfo{year}{2014}).
\newblock \eprint{arXiv:1411.4028}.

\bibitem{Guerreschi2019}
\bibinfo{author}{Guerreschi, G.~G.} \& \bibinfo{author}{Matsuura, A.~Y.}
\newblock \bibinfo{title}{{QAOA for Max-Cut requires hundreds of qubits for
  quantum speed-up}}.
\newblock \emph{\bibinfo{journal}{Sci. Rep.}} \textbf{\bibinfo{volume}{9}},
  \bibinfo{pages}{6903} (\bibinfo{year}{2019}).
\newblock \urlprefix\url{https://doi.org/10.1038/s41598-019-43176-9}.

\bibitem{Shaydulin2019}
\bibinfo{author}{{Shaydulin}, R.} \emph{et~al.}
\newblock \bibinfo{title}{{A Hybrid Approach for Solving Optimization Problems
  on Small Quantum Computers}}.
\newblock \emph{\bibinfo{journal}{Computer}} \textbf{\bibinfo{volume}{52}},
  \bibinfo{pages}{18--26} (\bibinfo{year}{2019}).

\bibitem{Havlicek2019}
\bibinfo{author}{Havl{\'\i}{\v c}ek, V.} \emph{et~al.}
\newblock \bibinfo{title}{Supervised learning with quantum-enhanced feature
  spaces}.
\newblock \emph{\bibinfo{journal}{Nature}} \textbf{\bibinfo{volume}{567}},
  \bibinfo{pages}{209--212} (\bibinfo{year}{2019}).
\newblock \urlprefix\url{https://doi.org/10.1038/s41586-019-0980-2}.

\bibitem{Mitarai2018}
\bibinfo{author}{Mitarai, K.}, \bibinfo{author}{Negoro, M.},
  \bibinfo{author}{Kitagawa, M.} \& \bibinfo{author}{Fujii, K.}
\newblock \bibinfo{title}{Quantum circuit learning}.
\newblock \emph{\bibinfo{journal}{Phys. Rev. A}} \textbf{\bibinfo{volume}{98}},
  \bibinfo{pages}{032309} (\bibinfo{year}{2018}).
\newblock \urlprefix\url{https://link.aps.org/doi/10.1103/PhysRevA.98.032309}.

\bibitem{Schuld2019}
\bibinfo{author}{Schuld, M.} \& \bibinfo{author}{Killoran, N.}
\newblock \bibinfo{title}{{Quantum Machine Learning in Feature Hilbert
  Spaces}}.
\newblock \emph{\bibinfo{journal}{Phys. Rev. Lett.}}
  \textbf{\bibinfo{volume}{122}}, \bibinfo{pages}{040504}
  (\bibinfo{year}{2019}).
\newblock
  \urlprefix\url{https://link.aps.org/doi/10.1103/PhysRevLett.122.040504}.

\bibitem{McClean2018}
\bibinfo{author}{McClean, J.~R.}, \bibinfo{author}{Boixo, S.},
  \bibinfo{author}{Smelyanskiy, V.~N.}, \bibinfo{author}{Babbush, R.} \&
  \bibinfo{author}{Neven, H.}
\newblock \bibinfo{title}{Barren plateaus in quantum neural network training
  landscapes}.
\newblock \emph{\bibinfo{journal}{Nat. Commun.}} \textbf{\bibinfo{volume}{9}},
  \bibinfo{pages}{4812} (\bibinfo{year}{2018}).
\newblock \urlprefix\url{https://doi.org/10.1038/s41467-018-07090-4}.

\bibitem{Woerner2019}
\bibinfo{author}{Woerner, S.} \& \bibinfo{author}{Egger, D.~J.}
\newblock \bibinfo{title}{Quantum risk analysis}.
\newblock \emph{\bibinfo{journal}{npj Quantum Inf.}}
  \textbf{\bibinfo{volume}{5}}, \bibinfo{pages}{15} (\bibinfo{year}{2019}).
\newblock \urlprefix\url{https://doi.org/10.1038/s41534-019-0130-6}.

\bibitem{Stamatopoulos2019}
\bibinfo{author}{Stamatopoulos, N.} \emph{et~al.}
\newblock \bibinfo{title}{{Option Pricing using Quantum Computers}}
  (\bibinfo{year}{2019}).
\newblock \eprint{arXiv:1905.02666}.

\bibitem{Egger2019}
\bibinfo{author}{Egger, D.~J.}, \bibinfo{author}{Gutiérrez, R.~G.},
  \bibinfo{author}{Mestre, J.~C.} \& \bibinfo{author}{Woerner, S.}
\newblock \bibinfo{title}{{Credit Risk Analysis using Quantum Computers}}
  (\bibinfo{year}{2019}).
\newblock \eprint{arXiv:1907.03044}.

\bibitem{Moll2018}
\bibinfo{author}{Moll, N.} \emph{et~al.}
\newblock \bibinfo{title}{Quantum optimization using variational algorithms on
  near-term quantum devices}.
\newblock \emph{\bibinfo{journal}{Quantum Sci. Technol.}}
  \textbf{\bibinfo{volume}{3}}, \bibinfo{pages}{030503} (\bibinfo{year}{2018}).
\newblock \urlprefix\url{http://stacks.iop.org/2058-9565/3/i=3/a=030503}.

\bibitem{McArdle2018}
\bibinfo{author}{McArdle, S.}, \bibinfo{author}{Endo, S.},
  \bibinfo{author}{Aspuru-Guzik, A.}, \bibinfo{author}{Benjamin, S.} \&
  \bibinfo{author}{Yuan, X.}
\newblock \bibinfo{title}{Quantum computational chemistry}
  (\bibinfo{year}{2018}).
\newblock \eprint{arXiv:1808.10402}.

\bibitem{Wecker2015}
\bibinfo{author}{Wecker, D.}, \bibinfo{author}{Hastings, M.~B.} \&
  \bibinfo{author}{Troyer, M.}
\newblock \bibinfo{title}{Progress towards practical quantum variational
  algorithms}.
\newblock \emph{\bibinfo{journal}{Phys. Rev. A}} \textbf{\bibinfo{volume}{92}},
  \bibinfo{pages}{042303} (\bibinfo{year}{2015}).
\newblock \urlprefix\url{https://link.aps.org/doi/10.1103/PhysRevA.92.042303}.

\bibitem{Wang2019}
\bibinfo{author}{Wang, D.}, \bibinfo{author}{Higgott, O.} \&
  \bibinfo{author}{Brierley, S.}
\newblock \bibinfo{title}{{Accelerated Variational Quantum Eigensolver}}.
\newblock \emph{\bibinfo{journal}{Phys. Rev. Lett.}}
  \textbf{\bibinfo{volume}{122}}, \bibinfo{pages}{140504}
  (\bibinfo{year}{2019}).
\newblock
  \urlprefix\url{https://link.aps.org/doi/10.1103/PhysRevLett.122.140504}.

\bibitem{McClean2016}
\bibinfo{author}{McClean, J.~R.}, \bibinfo{author}{Romero, J.},
  \bibinfo{author}{Babbush, R.} \& \bibinfo{author}{Aspuru-Guzik, A.}
\newblock \bibinfo{title}{The theory of variational hybrid quantum-classical
  algorithms}.
\newblock \emph{\bibinfo{journal}{New J. Phys.}} \textbf{\bibinfo{volume}{18}},
  \bibinfo{pages}{023023} (\bibinfo{year}{2016}).
\newblock \urlprefix\url{http://stacks.iop.org/1367-2630/18/i=2/a=023023}.

\bibitem{Bravyi2017}
\bibinfo{author}{Bravyi, S.}, \bibinfo{author}{Gambetta, J.~M.},
  \bibinfo{author}{Mezzacapo, A.} \& \bibinfo{author}{Temme, K.}
\newblock \bibinfo{title}{Tapering off qubits to simulate fermionic
  hamiltonians} (\bibinfo{year}{2017}).
\newblock \eprint{arXiv:1701.08213}.

\bibitem{Hamamura2018}
\bibinfo{author}{Hamamura, I.}
\newblock \bibinfo{title}{Separability criterion for quantum effects}.
\newblock \emph{\bibinfo{journal}{Phys. Lett. A}}
  \textbf{\bibinfo{volume}{382}}, \bibinfo{pages}{2573 -- 2577}
  (\bibinfo{year}{2018}).
\newblock
  \urlprefix\url{http://www.sciencedirect.com/science/article/pii/S0375960118307163}.

\bibitem{Jordan1928}
\bibinfo{author}{Jordan, P.} \& \bibinfo{author}{Wigner, E.}
\newblock \bibinfo{title}{{{\"U}ber das Paulische {\"A}quivalenzverbot}}.
\newblock \emph{\bibinfo{journal}{Z. Phys.}} \textbf{\bibinfo{volume}{47}},
  \bibinfo{pages}{631--651} (\bibinfo{year}{1928}).
\newblock \urlprefix\url{https://doi.org/10.1007/BF01331938}.

\bibitem{Bravyi2002}
\bibinfo{author}{Bravyi, S.~B.} \& \bibinfo{author}{Kitaev, A.~Y.}
\newblock \bibinfo{title}{{Fermionic Quantum Computation}}.
\newblock \emph{\bibinfo{journal}{Ann. Phys.}} \textbf{\bibinfo{volume}{298}},
  \bibinfo{pages}{210 -- 226} (\bibinfo{year}{2002}).
\newblock
  \urlprefix\url{http://www.sciencedirect.com/science/article/pii/S0003491602962548}.

\bibitem{Seeley2012}
\bibinfo{author}{Seeley, J.~T.}, \bibinfo{author}{Richard, M.~J.} \&
  \bibinfo{author}{Love, P.~J.}
\newblock \bibinfo{title}{{The Bravyi-Kitaev transformation for quantum
  computation of electronic structure}}.
\newblock \emph{\bibinfo{journal}{J. Chem. Phys.}}
  \textbf{\bibinfo{volume}{137}}, \bibinfo{pages}{224109}
  (\bibinfo{year}{2012}).
\newblock \urlprefix\url{https://doi.org/10.1063/1.4768229}.
\newblock \eprint{https://doi.org/10.1063/1.4768229}.

\bibitem{Entanglion}
\bibinfo{title}{{Entanglion}}.
\newblock \bibinfo{howpublished}{\url{https://entanglion.github.io/}}
  (\bibinfo{year}{2018}).

\bibitem{Jena2019}
\bibinfo{author}{Jena, A.}, \bibinfo{author}{Genin, S.} \&
  \bibinfo{author}{Mosca, M.}
\newblock \bibinfo{title}{{Pauli Partitioning with Respect to Gate Sets}}
  (\bibinfo{year}{2019}).
\newblock \eprint{arXiv:1907.07859}.

\bibitem{Yen2019}
\bibinfo{author}{Yen, T.-C.}, \bibinfo{author}{Verteletskyi, V.} \&
  \bibinfo{author}{Izmaylov, A.~F.}
\newblock \bibinfo{title}{Measuring all compatible operators in one series of a
  single-qubit measurements using unitary transformations}
  (\bibinfo{year}{2019}).
\newblock \eprint{arXiv:1907.09386}.

\bibitem{Gokhale2019}
\bibinfo{author}{Gokhale, P.} \emph{et~al.}
\newblock \bibinfo{title}{{Minimizing State Preparations in Variational Quantum
  Eigensolver by Partitioning into Commuting Families}} (\bibinfo{year}{2019}).
\newblock \eprint{arXiv:1907.13623}.

\bibitem{Konc2007}
\bibinfo{author}{Konc, J.} \& \bibinfo{author}{Jane\v{z}i\v{c}, D.}
\newblock \bibinfo{title}{An improved branch and bound algorithm for the
  maximum clique problem}.
\newblock \emph{\bibinfo{journal}{MATCH Commun. Math. Comput. Chem.}}
  \textbf{\bibinfo{volume}{58}}, \bibinfo{pages}{569--590}
  (\bibinfo{year}{2007}).
\newblock \urlprefix\url{http://match.pmf.kg.ac.rs/content58n3.htm}.

\bibitem{Schrijver2003}
\bibinfo{author}{Schrijver, A.}
\newblock \emph{\bibinfo{title}{Combinatorial optimization: polyhedra and
  efficiency}} (\bibinfo{publisher}{Springer}, \bibinfo{year}{2003}).

\bibitem{Temme2017}
\bibinfo{author}{Temme, K.}, \bibinfo{author}{Bravyi, S.} \&
  \bibinfo{author}{Gambetta, J.~M.}
\newblock \bibinfo{title}{{Error Mitigation for Short-Depth Quantum Circuits}}.
\newblock \emph{\bibinfo{journal}{Phys. Rev. Lett.}}
  \textbf{\bibinfo{volume}{119}}, \bibinfo{pages}{180509}
  (\bibinfo{year}{2017}).
\newblock
  \urlprefix\url{https://link.aps.org/doi/10.1103/PhysRevLett.119.180509}.

\bibitem{Corcoles2019}
\bibinfo{author}{Corcoles, A.~D.} \emph{et~al.}
\newblock \bibinfo{title}{{Challenges and Opportunities of Near-Term Quantum
  Computing Systems}} (\bibinfo{year}{2019}).
\newblock \eprint{arXiv:1910.02894}.

\bibitem{Spall1992}
\bibinfo{author}{{Spall}, J.~C.}
\newblock \bibinfo{title}{Multivariate stochastic approximation using a
  simultaneous perturbation gradient approximation}.
\newblock \emph{\bibinfo{journal}{IEEE Trans. Autom. Control}}
  \textbf{\bibinfo{volume}{37}}, \bibinfo{pages}{332--341}
  (\bibinfo{year}{1992}).

\bibitem{Chow2010}
\bibinfo{author}{Chow, J.~M.} \emph{et~al.}
\newblock \bibinfo{title}{Detecting highly entangled states with a joint qubit
  readout}.
\newblock \emph{\bibinfo{journal}{Phys. Rev. A}} \textbf{\bibinfo{volume}{81}},
  \bibinfo{pages}{062325} (\bibinfo{year}{2010}).
\newblock \urlprefix\url{https://link.aps.org/doi/10.1103/PhysRevA.81.062325}.

\bibitem{Izmaylov2019}
\bibinfo{author}{Izmaylov, A.~F.}, \bibinfo{author}{Yen, T.-C.} \&
  \bibinfo{author}{Ryabinkin, I.~G.}
\newblock \bibinfo{title}{Revising the measurement process in the variational
  quantum eigensolver: is it possible to reduce the number of separately
  measured operators?}
\newblock \emph{\bibinfo{journal}{Chem. Sci.}} \textbf{\bibinfo{volume}{10}},
  \bibinfo{pages}{3746--3755} (\bibinfo{year}{2019}).
\newblock \urlprefix\url{http://dx.doi.org/10.1039/C8SC05592K}.

\bibitem{Nakanishi2019}
\bibinfo{author}{Nakanishi, K.~M.}, \bibinfo{author}{Fujii, K.} \&
  \bibinfo{author}{Todo, S.}
\newblock \bibinfo{title}{Sequential minimal optimization for quantum-classical
  hybrid algorithms} (\bibinfo{year}{2019}).
\newblock \eprint{arXiv:1903.12166}.

\bibitem{Rubin2018}
\bibinfo{author}{Rubin, N.~C.}, \bibinfo{author}{Babbush, R.} \&
  \bibinfo{author}{McClean, J.}
\newblock \bibinfo{title}{Application of fermionic marginal constraints to
  hybrid quantum algorithms}.
\newblock \emph{\bibinfo{journal}{New J. Phys.}} \textbf{\bibinfo{volume}{20}},
  \bibinfo{pages}{053020} (\bibinfo{year}{2018}).
\newblock \urlprefix\url{https://doi.org/10.1088%2F1367-2630%2Faab919}.

\bibitem{Setia2018}
\bibinfo{author}{Setia, K.} \& \bibinfo{author}{Whitfield, J.~D.}
\newblock \bibinfo{title}{{Bravyi-Kitaev Superfast simulation of electronic
  structure on a quantum computer}}.
\newblock \emph{\bibinfo{journal}{J. Chem. Phys.}}
  \textbf{\bibinfo{volume}{148}}, \bibinfo{pages}{164104}
  (\bibinfo{year}{2018}).
\newblock \urlprefix\url{https://doi.org/10.1063/1.5019371}.
\newblock \eprint{https://doi.org/10.1063/1.5019371}.

\bibitem{Babbush2018}
\bibinfo{author}{Babbush, R.} \emph{et~al.}
\newblock \bibinfo{title}{{Low-Depth Quantum Simulation of Materials}}.
\newblock \emph{\bibinfo{journal}{Phys. Rev. X}} \textbf{\bibinfo{volume}{8}},
  \bibinfo{pages}{011044} (\bibinfo{year}{2018}).
\newblock \urlprefix\url{https://link.aps.org/doi/10.1103/PhysRevX.8.011044}.

\bibitem{Barkoutsos2018}
\bibinfo{author}{Barkoutsos, P.~K.} \emph{et~al.}
\newblock \bibinfo{title}{Quantum algorithms for electronic structure
  calculations: Particle-hole hamiltonian and optimized wave-function
  expansions}.
\newblock \emph{\bibinfo{journal}{Phys. Rev. A}} \textbf{\bibinfo{volume}{98}},
  \bibinfo{pages}{022322} (\bibinfo{year}{2018}).
\newblock \urlprefix\url{https://link.aps.org/doi/10.1103/PhysRevA.98.022322}.

\bibitem{Romero2018}
\bibinfo{author}{Romero, J.} \emph{et~al.}
\newblock \bibinfo{title}{Strategies for quantum computing molecular energies
  using the unitary coupled cluster ansatz}.
\newblock \emph{\bibinfo{journal}{Quantum Sci. Technol.}}
  \textbf{\bibinfo{volume}{4}}, \bibinfo{pages}{014008} (\bibinfo{year}{2018}).
\newblock \urlprefix\url{https://doi.org/10.1088%2F2058-9565%2Faad3e4}.

\bibitem{Gard2019}
\bibinfo{author}{Gard, B.~T.} \emph{et~al.}
\newblock \bibinfo{title}{{Efficient Symmetry-Preserving State Preparation
  Circuits for the Variational Quantum Eigensolver Algorithm}}
  (\bibinfo{year}{2019}).
\newblock \eprint{arXiv:1904.10910}.

\bibitem{OpenFermion}
\bibinfo{author}{McClean, J.~R.} \emph{et~al.}
\newblock \bibinfo{title}{{OpenFermion: The Electronic Structure Package for
  Quantum Computers}} (\bibinfo{year}{2017}).
\newblock \eprint{arXiv:1710.07629}.

\bibitem{Verteletskyi2019}
\bibinfo{author}{Verteletskyi, V.}, \bibinfo{author}{Yen, T.-C.} \&
  \bibinfo{author}{Izmaylov, A.~F.}
\newblock \bibinfo{title}{{Measurement Optimization in the Variational Quantum
  Eigensolver Using a Minimum Clique Cover}} (\bibinfo{year}{2019}).
\newblock \eprint{arXiv:1907.03358}.

\bibitem{Izmaylov2019b}
\bibinfo{author}{Izmaylov, A.~F.}, \bibinfo{author}{Yen, T.-C.},
  \bibinfo{author}{Lang, R.~A.} \& \bibinfo{author}{Verteletskyi, V.}
\newblock \bibinfo{title}{{Unitary partitioning approach to the measurement
  problem in the Variational Quantum Eigensolver method}}
  (\bibinfo{year}{2019}).
\newblock \eprint{arXiv:1907.09040}.

\bibitem{Huggins2019}
\bibinfo{author}{Huggins, W.~J.} \emph{et~al.}
\newblock \bibinfo{title}{{Efficient and Noise Resilient Measurements for
  Quantum Chemistry on Near-Term Quantum Computers}} (\bibinfo{year}{2019}).
\newblock \eprint{arXiv:1907.13117}.

\bibitem{Crawford2019}
\bibinfo{author}{Crawford, O.} \emph{et~al.}
\newblock \bibinfo{title}{{Efficient quantum measurement of Pauli operators}}
  (\bibinfo{year}{2019}).
\newblock \eprint{arXiv:1908.06942}.

\bibitem{Zhao2019}
\bibinfo{author}{Zhao, A.} \emph{et~al.}
\newblock \bibinfo{title}{Measurement reduction in variational quantum
  algorithms} (\bibinfo{year}{2019}).
\newblock \eprint{arXiv:1908.08067}.

\bibitem{Gokhale2019b}
\bibinfo{author}{Gokhale, P.} \& \bibinfo{author}{Chong, F.~T.}
\newblock \bibinfo{title}{{$O(N^3)$ Measurement Cost for Variational Quantum
  Eigensolver on Molecular Hamiltonians}} (\bibinfo{year}{2019}).
\newblock \eprint{arXiv:1908.11857}.

\bibitem{Jiang2019}
\bibinfo{author}{Jiang, Z.}, \bibinfo{author}{Kalev, A.},
  \bibinfo{author}{Mruczkiewicz, W.} \& \bibinfo{author}{Neven, H.}
\newblock \bibinfo{title}{Optimal fermion-to-qubit mapping via ternary trees
  with applications to reduced quantum states learning} (\bibinfo{year}{2019}).
\newblock \eprint{arXiv:1910.10746}.

\bibitem{Bell1964}
\bibinfo{author}{Bell, J.~S.}
\newblock \bibinfo{title}{{On the Einstein Podolsky Rosen Paradox*}}.
\newblock \emph{\bibinfo{journal}{Physics}} \textbf{\bibinfo{volume}{1}},
  \bibinfo{pages}{195--290} (\bibinfo{year}{1964}).

\bibitem{Clauser1969}
\bibinfo{author}{Clauser, J.~F.}, \bibinfo{author}{Horne, M.~A.},
  \bibinfo{author}{Shimony, A.} \& \bibinfo{author}{Holt, R.~A.}
\newblock \bibinfo{title}{{Proposed Experiment to Test Local Hidden-Variable
  Theories}}.
\newblock \emph{\bibinfo{journal}{Phys. Rev. Lett.}}
  \textbf{\bibinfo{volume}{23}}, \bibinfo{pages}{880--884}
  (\bibinfo{year}{1969}).

\bibitem{Bennett1993}
\bibinfo{author}{Bennett, C.~H.} \emph{et~al.}
\newblock \bibinfo{title}{{Teleporting an unknown quantum state via dual
  classical and Einstein-Podolsky-Rosen channels}}.
\newblock \emph{\bibinfo{journal}{Phys. Rev. Lett.}}
  \textbf{\bibinfo{volume}{70}}, \bibinfo{pages}{1895--1899}
  (\bibinfo{year}{1993}).
\newblock \urlprefix\url{https://link.aps.org/doi/10.1103/PhysRevLett.70.1895}.

\end{thebibliography}

\clearpage
\widetext
\setcounter{equation}{0}
\setcounter{figure}{0}
\setcounter{table}{0}
\setcounter{page}{1}

\renewcommand{\theequation}{S\arabic{equation}}
\renewcommand{\thefigure}{S\arabic{figure}}
\begin{center}
  \textbf{\Large Supplementary Information for ``Efficient evaluation of quantum observables using entangled measurements''}
\end{center}

\appendix

\section{Pauli graph and LDFC algorithm}\label{sec:pauli}

In this section, we explain a previous method called grouping with TPB in the main text.
The grouping algorithm consists of two parts: the construction of the Pauli graph and its coloring algorithm called the largest degree first coloring.
These algorithms were implemented in the Qiskit~\cite{Qiskit}.

The Pauli graph represents the noncommutativity of the Pauli strings.
Its nodes are Pauli strings, and its edges exist if and only if two nodes are not qubit-wise commutative.
The algorithm which constructs the Pauli graph using TPB is shown in Algorithm~\ref{alg:pauli-graph}.
\begin{algorithm}[H]
	\caption{The construction of Pauli graph}
	\label{alg:pauli-graph}
	\begin{algorithmic}[1]
		\STATE \textbf{Input}: observable $A = \sum_{i=1}^n a_i P_i$.  Note that each Pauli string has length $N$.
    \STATE Make graph $G=(V,E)$ with $n$ nodes and no edges.
    \FOR{$i = 1,\dots n-1$}
      \FOR{$j = i+1,\dots n$}
        \IF{Any one of the Pauli operators of Pauli strings $P_i$ and $P_j$ are noncommutative}
          \STATE Span an edge between $i$-th node and $j$-th node.
        \ENDIF
      \ENDFOR
    \ENDFOR
		\STATE Return the graph $G$.
	\end{algorithmic}
\end{algorithm}

After the construction of the Pauli graph, we label the nodes of the graph with the rule that the neighbor nodes have different labels.
The label is called color.
Since the coloring problem is hard to compute, we use the heuristic algorithm, such as the largest degree first coloring.
We explain the algorithm in Algorithm~\ref{alg:ldfc}.

\begin{algorithm}[H]
  \caption{The largest degree first coloring}
	\label{alg:ldfc}
	\begin{algorithmic}[1]
    \STATE \textbf{Input}: graph $G=(V,E)$.
    \STATE Sort the nodes in the descending order of degree on G and let them be $V=\set{v_1,\dots,v_n }$.
    \STATE Initialize a list of colors $C=\set{c_1,\dots,c_n}$.
    Elements of $C$ are integer numbers representing colors.
    Let unassigned color be $-1$.
    Initialization sets all elements to $-1$.
    \FOR{$i = 1,\dots,n$}
      \STATE Asign the minimum color number which is not assigned to neighbors of $v_i$ to $c_i$.
    \ENDFOR
		\STATE Return the list of colors $C$.
	\end{algorithmic}
\end{algorithm}

\section{Examples of entangled measurements}\label{sec:exa-ent}

\begin{figure}[b]
  \centering
  \includegraphics[width=0.5\linewidth]{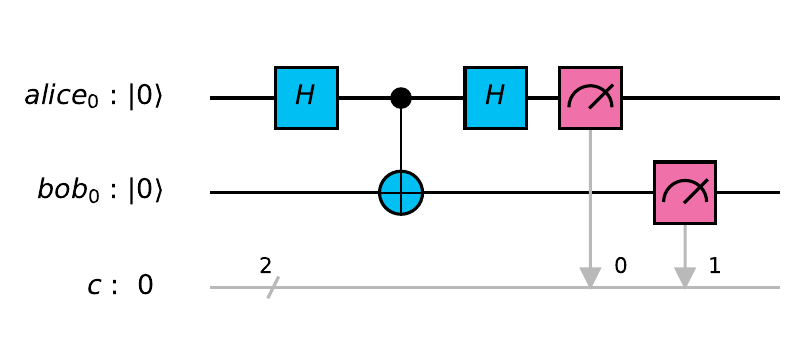}
  \caption{%
  Quantum circuit of an omega-YY measurement.
}
\label{fig:omega-y}
\end{figure}

The first example of entangled measurements, Bell measurements, is shown in the Method section of the main text.
The second example of entangled measurements is an omega measurement that is a projective measurement on omega states introduced in the board game Entanglion developed by IBM Research~\cite{Entanglion}.
Omega states are defined as follows:
\begin{align}
  \Ket{\Omega^{YY}_0} &= \half\ket{00} + \half\ket{01} - \half\ket{10} + \half\ket{11}, \\
  \Ket{\Omega^{YY}_1} &= -\half\ket{00} + \half\ket{01} + \half\ket{10} + \half\ket{11}, \\
  \Ket{\Omega^{YY}_2} &= \half\ket{00} + \half\ket{01} + \half\ket{10} - \half\ket{11}, \\
  \Ket{\Omega^{YY}_3} &= \half\ket{00} - \half\ket{01} + \half\ket{10} + \half\ket{11}.
\end{align}
Herein, we note that omega states are described as omega-YY states because other omega states that can be introduced subsequently.
The expectation values of $\Y\otimes\Y$, $\X\otimes\Z$, and $\Z\otimes\X$ can be obtained from an omega-YY measurement
since the following identities hold:
\begin{align}
  \Y\otimes\Y &= \Ketbra{\Omega^{YY}_1} + \Ketbra{\Omega^{YY}_2} - \Ketbra{\Omega^{YY}_0} - \Ketbra{\Omega^{YY}_3}, \\
  \X\otimes\Z &= \Ketbra{\Omega^{YY}_2} + \Ketbra{\Omega^{YY}_3} - \Ketbra{\Omega^{YY}_0} - \Ketbra{\Omega^{YY}_1}, \\
  \Z\otimes\X &= \Ketbra{\Omega^{YY}_0} + \Ketbra{\Omega^{YY}_2} - \Ketbra{\Omega^{YY}_1} - \Ketbra{\Omega^{YY}_3}.
\end{align}
Circuit implementation of an omega-YY measurement is shown in Fig.~\ref{fig:omega-y}.

\begin{figure}
  \begin{subfigure}[b]{.5\linewidth}
    \centering
    \includegraphics[height=10em]{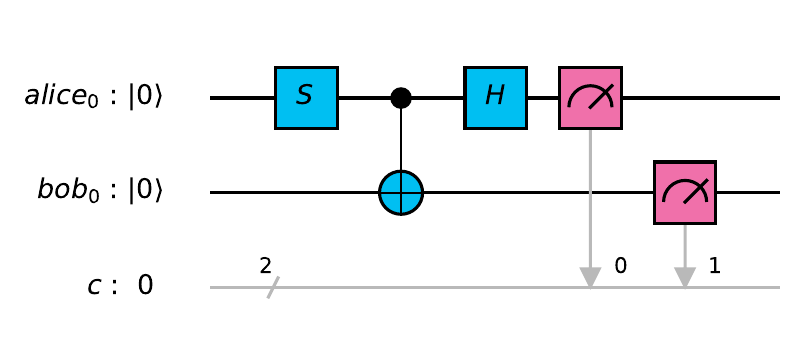}
    \caption{Omega-ZZ measurement.}
    \label{fig:omega-z}
  \end{subfigure}%
  \begin{subfigure}[b]{.5\linewidth}
    \centering
    \includegraphics[height=10em]{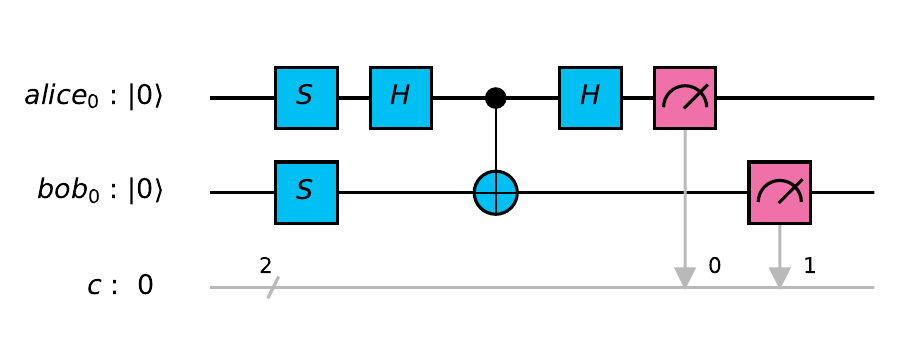}
    \caption{Omega-XX measurement.}
    \label{fig:omega-x}
  \end{subfigure}
  \caption{Quantum circuits of omega-ZZ and omega-XX measurements.}
  \label{fig:omega-zx}
\end{figure}

The third and fourth examples are an omega-ZZ measurement and an omega-XX measurement that are projective measurements on the following states:
\begin{align}
  \Ket{\Omega^{ZZ}_0} &= \sqrthalf \ket{01} - i \sqrthalf\ket{10}, \\
  \Ket{\Omega^{ZZ}_1} &= \sqrthalf \ket{01} + i \sqrthalf\ket{10}, \\
  \Ket{\Omega^{ZZ}_2} &= \sqrthalf \ket{00} - i \sqrthalf\ket{11}, \\
  \Ket{\Omega^{ZZ}_3} &= \sqrthalf \ket{00} + i \sqrthalf\ket{11},
\end{align}
and
\begin{align}
  \Ket{\Omega^{XX}_0} &= \half\ket{00} - \frac{i}{2}\ket{01} + \frac{i}{2}\ket{10} - \half\ket{11}, \\
  \Ket{\Omega^{XX}_1} &= -\half\ket{00} - \frac{i}{2}\ket{01} - \frac{i}{2}\ket{10} - \half\ket{11}, \\
  \Ket{\Omega^{XX}_2} &= \half\ket{00} - \frac{i}{2}\ket{01} - \frac{i}{2}\ket{10} + \half\ket{11}, \\
  \Ket{\Omega^{XX}_3} &= \half\ket{00} + \frac{i}{2}\ket{01} - \frac{i}{2}\ket{10} - \half\ket{11}.
\end{align}
In the same way as a Bell measurement and an omega-YY measurement, the following two states allow joint measurements of the two-qubit operators.
An omega-ZZ measurement is a joint measurement of observables $\Z\otimes\Z$, $\X\otimes\X$, and $\Y\otimes\X$.
It can be derived from the following identities:
\begin{align}
  \Z\otimes\Z &= \Ketbra{\Omega^{ZZ}_2} + \Ketbra{\Omega^{ZZ}_3} - \Ketbra{\Omega^{ZZ}_0} - \Ketbra{\Omega^{ZZ}_1}, \\
  \X\otimes\Y &= \Ketbra{\Omega^{ZZ}_0} + \Ketbra{\Omega^{ZZ}_3} - \Ketbra{\Omega^{ZZ}_1} - \Ketbra{\Omega^{ZZ}_2}, \\
  \Y\otimes\X &= \Ketbra{\Omega^{ZZ}_1} + \Ketbra{\Omega^{ZZ}_3} - \Ketbra{\Omega^{ZZ}_0} - \Ketbra{\Omega^{ZZ}_2}.
\end{align}
An omega-XX measurement is a joint measurement of observables $\X\otimes\X$, $\Y\otimes\Z$, and $\Z\otimes\Y$.
It can be derived from the following identities:
\begin{align}
  \X\otimes\X &= \Ketbra{\Omega^{XX}_1} + \Ketbra{\Omega^{XX}_2} - \Ketbra{\Omega^{XX}_0} - \Ketbra{\Omega^{XX}_3}, \\
  \Y\otimes\Z &= \Ketbra{\Omega^{XX}_0} + \Ketbra{\Omega^{XX}_1} - \Ketbra{\Omega^{XX}_2} - \Ketbra{\Omega^{XX}_3}, \\
  \Z\otimes\Y &= \Ketbra{\Omega^{XX}_1} + \Ketbra{\Omega^{XX}_3} - \Ketbra{\Omega^{XX}_0} - \Ketbra{\Omega^{XX}_2}.
\end{align}
Circuit implementations of an omega-ZZ measurement and an omega-XX measurement are shown in Fig.~\ref{fig:omega-zx}.

\begin{figure}
  \centering
  \includegraphics[height=10em]{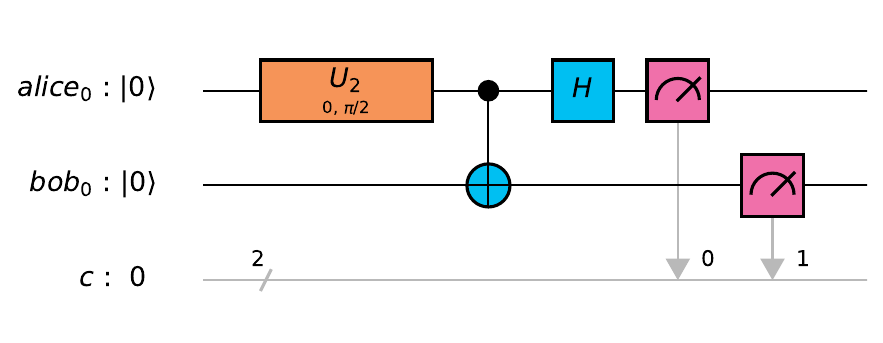}
  \caption{%
  Quantum circuit of a chi measurement.
}
\label{fig:chi}
\end{figure}

The fifth example is a chi measurement.
A chi measurement is a joint measurement of $\X\otimes\Y$, $\Y\otimes\Z$, and $\Z\otimes\X$.
Let us define the following states:
\begin{align}
  \Ket{X_0} &= \half\ket{00} + \half\ket{01} + \frac{i}{2}\ket{10} - \frac{i}{2}\ket{11}, \\
  \Ket{X_1} &= \half\ket{00} + \half\ket{01} - \frac{i}{2}\ket{10} + \frac{i}{2}\ket{11}, \\
  \Ket{X_2} &= \half\ket{00} - \half\ket{01} + \frac{i}{2}\ket{10} + \frac{i}{2}\ket{11}, \\
  \Ket{X_3} &= - \half\ket{00} + \half\ket{01} + \frac{i}{2}\ket{10} + \frac{i}{2}\ket{11}.
\end{align}
A chi measurement is a projective measurement of these states as the following identities hold:
\begin{align}
  \X\otimes\Y &= -\Ketbra{X_0}+\Ketbra{X_1}+\Ketbra{X_2}-\Ketbra{X_3}, \\
  \Y\otimes\Z &= \Ketbra{X_0}-\Ketbra{X_1}+\Ketbra{X_2}-\Ketbra{X_3},\\
  \Z\otimes\X &= \Ketbra{X_0}+\Ketbra{X_1}-\Ketbra{X_2}-\Ketbra{X_3}.
\end{align}
The circuit implementation of a chi measurement is shown in Fig.~\ref{fig:chi}.
Herein, we use a one-pulse gate $\mathrm{U_2}$ defined in Qiskit Terra~\cite{Qiskit}.
$\mathrm{U_2}$-gate has two parameters $\phi$ and $\lambda$.
The matrix representation of $\mathrm{U_2}$-gate is
\begin{equation}
  \sqrthalf \begin{pmatrix}
    1 & -e^{i\lambda} \\
    e^{i\phi} & e^{i(\phi+\lambda)}
  \end{pmatrix}
\end{equation}
In Fig.~\ref{fig:chi}, the parameters of $\mathrm{U_2}$-gate can be expressed as $\phi=0$ and $\lambda=\frac{\pi}{2}$.

There are entangled measurements with three or more qubits for joint measurement of Pauli strings
and they can be realized as similar formulations;
however, we used only two-qubit entangled measurements in our experiments.
This is because large quantum circuits are required for multipartite entangled measurements in general
and it seems difficult to execute such circuits on current quantum computers owing to possible errors that may occur.

\section{Error derived from the grouping}\label{sec:error}
\subsection{Theoretical analysis of the covariance effects by grouping}\label{sec:varthm}

We assume that observables are given by the following weighted sum of Pauli strings:
\begin{equation}
  A = \sum_{i=1}^n a_i P_i
\end{equation}
where $a_i$ denotes a real number and $P_i$ denotes a Pauli string.
In this section, let us consider the standard error derived from the grouping of Pauli strings.
Let a grouping divide $n$ Pauli strings into $K$ sets $s_1, s_2, \dots, s_K$.
Then, the equality
\begin{equation}
  A = \sum_{k=1}^{K} \sum_{i\in s_k} a_i P_i,
\end{equation}
holds by definition.
If the number of samples for group $s_k$ is $S_k$, the standard error can be expressed as
\begin{align}
  \epsilon_G &= \sqrt{\sum_{k=1}^{K} \frac{\Var{\sum_{i\in s_k}a_i P_i}}{S_k}}\\
  &= \sqrt{\sum_{k=1}^{K} \frac{\sum_{(i,j)\in s_k\times s_k} a_i a_j \Braket{(P_i-\Braket{P_i})(P_j-\Braket{P_j})}}{S_k}}.
\end{align}
In the case of no-grouping, i.e., $|s_k|=1$, the standard error can be expressed as
\begin{equation}
  \epsilon_{NG} = \sqrt{\frac{1}{S}\sum_{i=1}^{n} a_i^2\Braket{{(P_i-\Braket{P_i})}^2}},
\end{equation}
where we assume that $S_k=S$ for simplicity using a constant positive integer $S$.
This error does not contain the covariance effects.

We propose that the number of samples in each group is proportional to the size of the group, i.e., $S_k = |s_k| S$,
where the total numbers of samples for no-grouping and grouping are the same.
Let us compare the standard error without grouping $\epsilon_{NG}$ and that with grouping $\epsilon_G$ as follows:
\begin{align}
  \epsilon_G^2
  &= \sum_{k=1}^{K} \frac{\sum_{(i,j)\in s_k\times s_k}a_i a_j \Braket{(P_i-\Braket{P_i})(P_j-\Braket{P_j})}}{S_k}\\
  &= \frac{1}{S}\sum_{k=1}^{K} \frac{\sum_{(i,j)\in s_k\times s_k}\Braket{(a_i P_i-\Braket{a_i P_i})(a_j P_j-\Braket{a_j P_j})}}{|s_k|}\\
  &\leq \frac{1}{S}\sum_{k=1}^{K} \frac{\sum_{(i,j)\in s_k\times s_k}\sqrt{\Braket{{(a_i P_i-\Braket{a_i P_i})}^2}\Braket{{(a_j P_j-\Braket{a_j P_j})}^2}}}{|s_k|} \label{eq:first}\\
  &\leq \frac{1}{S}\sum_{k=1}^{K} \frac{\sum_{(i,j)\in s_k\times s_k}\left[\Braket{{(a_i P_i-\Braket{a_i P_i})}^2}+\Braket{{(a_j P_j-\Braket{a_j P_j})}^2}\right]}{2|s_k|} \label{eq:second}\\
  &= \frac{1}{S}\sum_{k=1}^{K} \sum_{i\in s_k}\Braket{{(a_i P_i-\Braket{a_i P_i})}^2} \\
  &= \epsilon_{NG}^2,
\end{align}
where we applied the Cauchy--Schwarz inequality to the first inequality~\eqref{eq:first} and the AM--GM inequality to the second inequality~\eqref{eq:second}.
As a result, we obtain the following theorem:
\begin{thm}\label{thm:grouping}
  The standard error of grouping is less or equal to that of no-grouping; i.e., the inequality
\begin{equation}
  \epsilon_G \leq \epsilon_{NG},
\end{equation}
holds.
\end{thm}

\subsection{Numerical experiment}\label{sec:num}

\begin{figure}[ht]
  \centering
  \begin{subfigure}[t]{.48\linewidth}
    \includegraphics[width=1\linewidth]{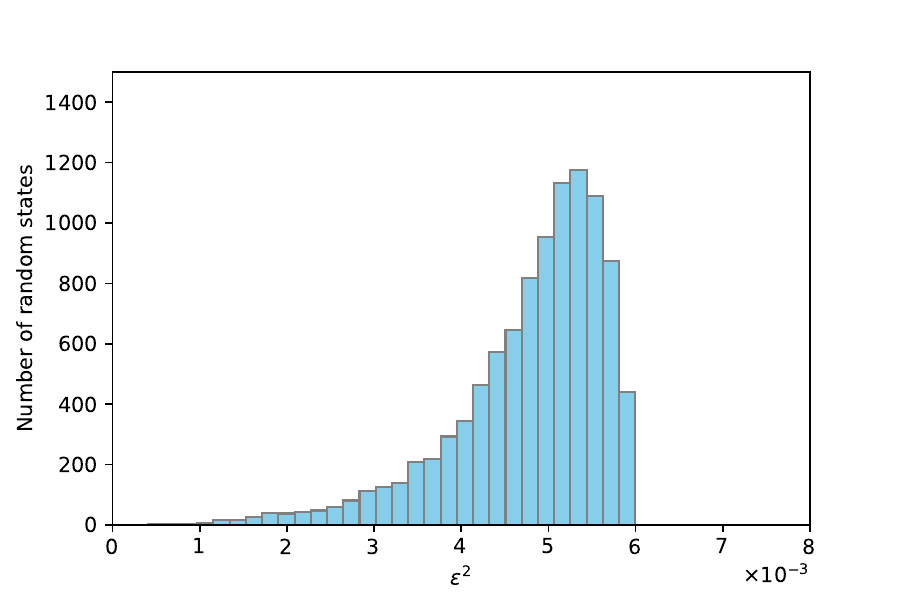}
    \caption{The square of the standard error for Pauli strings (no-grouping) is $\Var{H}_{NG}/500$. We run 500 samples per each of measurements.}
    \label{fig:ug}
  \end{subfigure}
  \hfill
  \begin{subfigure}[t]{.48\linewidth}
    \centering
    \includegraphics[width=1\linewidth]{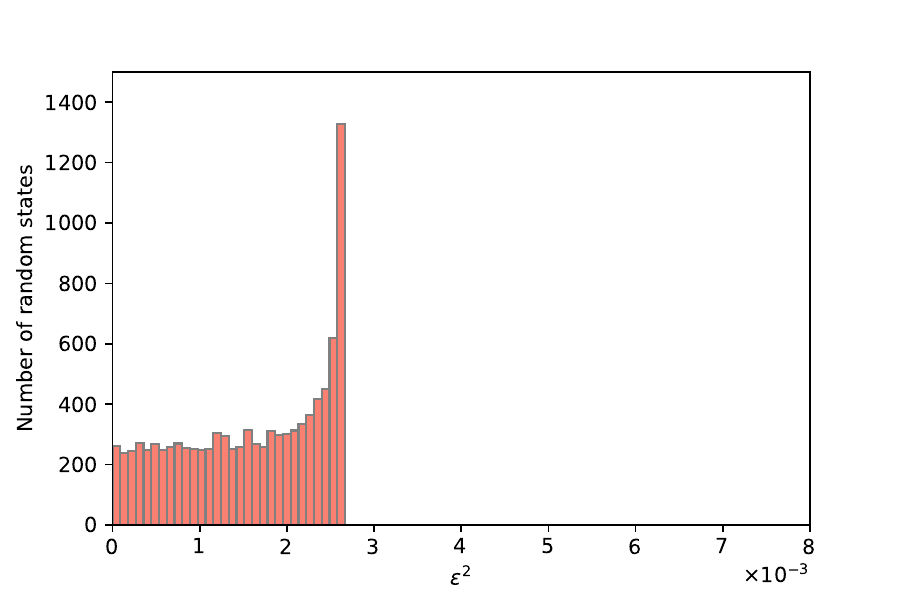}
    \caption{The square of the standard error for Pauli strings (grouping by a Bell measurement) is $\Var{H}_G/1500$. We run 1500 samples of the Bell measurement.}
    \label{fig:g}
  \end{subfigure}
  \caption{%
  Comparison of distributions of the square of standard errors.
  Both values are computed for 10,000 random states.
  From Theorem~\ref{thm:grouping}, the values on the left are greater than those on the right for the same quantum state.
  }
  \label{fig:error}
\end{figure}

Herein, we present the two-qubit antiferromagnetic Heisenberg model,
which can be expressed as
\begin{equation}
  H = J(\X\X + \Y\Y + \Z\Z),
\end{equation}
with $J > 0$.
We assume $J=1$ for simplicity.
A variance of this Hamiltonian is
\begin{equation}
  \Var{H}_{NG} = \Var{\X\X} + \Var{\Y\Y} + \Var{\Z\Z},
\end{equation}
if one performs measurements $\X\X$, $\Y\Y$, and $\Z\Z$ independently.
If one performs a joint measurement (Bell measurement), a variance of the Hamiltonian can then be expressed as
\begin{align*}
  \Var{H}_G =& \Var{\X\X+\Y\Y+\Z\Z} \\
            =& \Var{\X\X} + \Var{\Y\Y} + \Var{\Z\Z} + 2\Cov{\X\X, \Y\Y}
            + 2\Cov{\Y\Y, \Z\Z} + 2\Cov{\Z\Z, \X\X}.
\end{align*}
We numerically computed the square of the standard errors for Pauli strings and grouped Pauli strings using a Bell measurement.
We used 10,000 random states of Haar random unitary to calculate the standard errors.
The number of running circuits is 500 samples per Pauli strings for no-grouping (three groups) and 1500 samples per group for a Bell measurement (one group).
The average of the squared standard errors of no-grouping is 0.00479
and that of the grouping with a Bell measurement is 0.00159.
This numerical result shows that the grouping with Bell measurement has less standard error.
See Fig.~\ref{fig:error} for the distributions.
The distribution for the grouping with a Bell measurement is interesting because the values are concentrated around the upper bound,
and this characteristic is not seen in the case of the grouping with TPB (refer to Fig.~S5.\ in the Supplementary Information of Kandala et~al.~\cite{Kandala2017}).
This might be an intrinsic property of the grouping with entangled measurements.

\section{Convergence of VQE on real quantum computers}\label{sec:failure}

\begin{figure}
  \centering
  \begin{subfigure}[b]{.49\linewidth}
    \includegraphics[width=.99\linewidth]{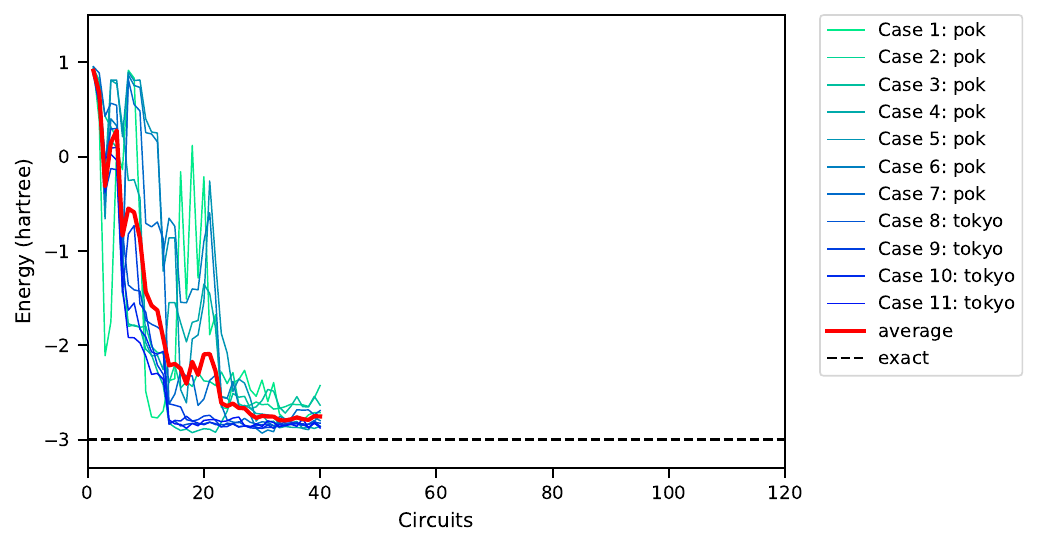}
    \caption{Grouping with Bell measurement with error mitigation.}
  \end{subfigure}
  \begin{subfigure}[b]{.49\linewidth}
    \includegraphics[width=.99\linewidth]{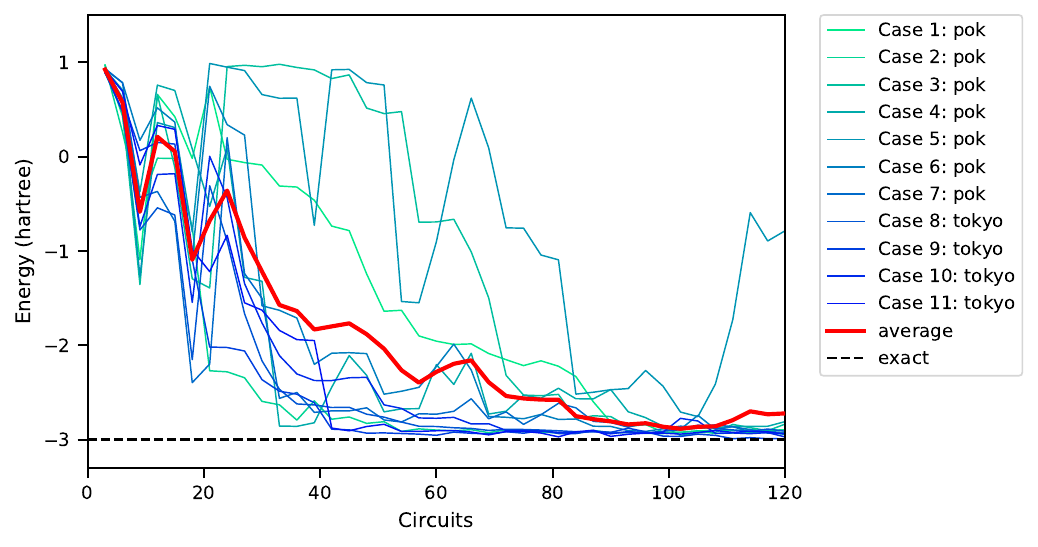}
    \caption{No-grouping with error mitigation}
  \end{subfigure}
  \begin{subfigure}[b]{.49\linewidth}
    \includegraphics[width=.99\linewidth]{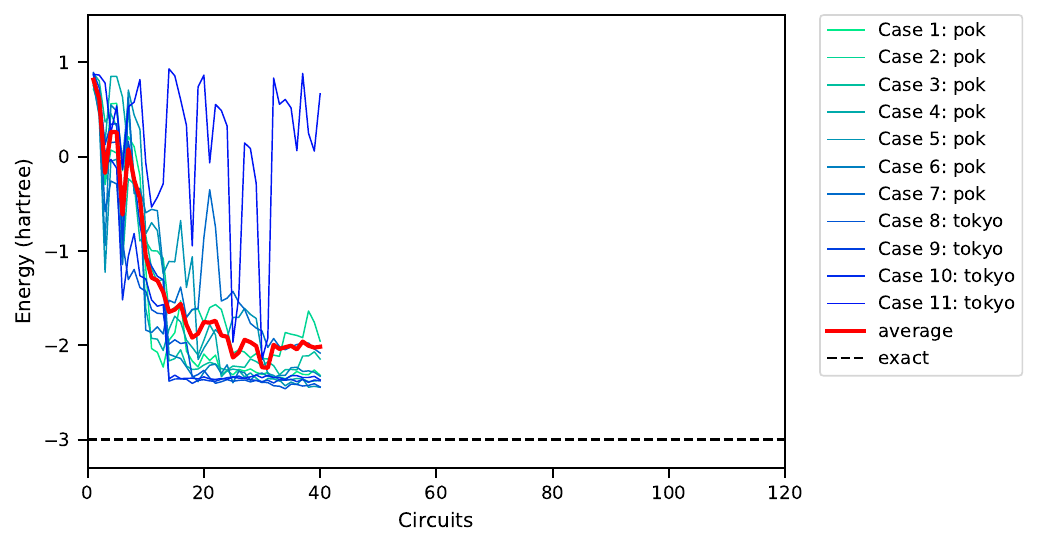}
    \caption{Grouping with Bell measurement without error mitigation.}
  \end{subfigure}
  \begin{subfigure}[b]{.49\linewidth}
    \includegraphics[width=.99\linewidth]{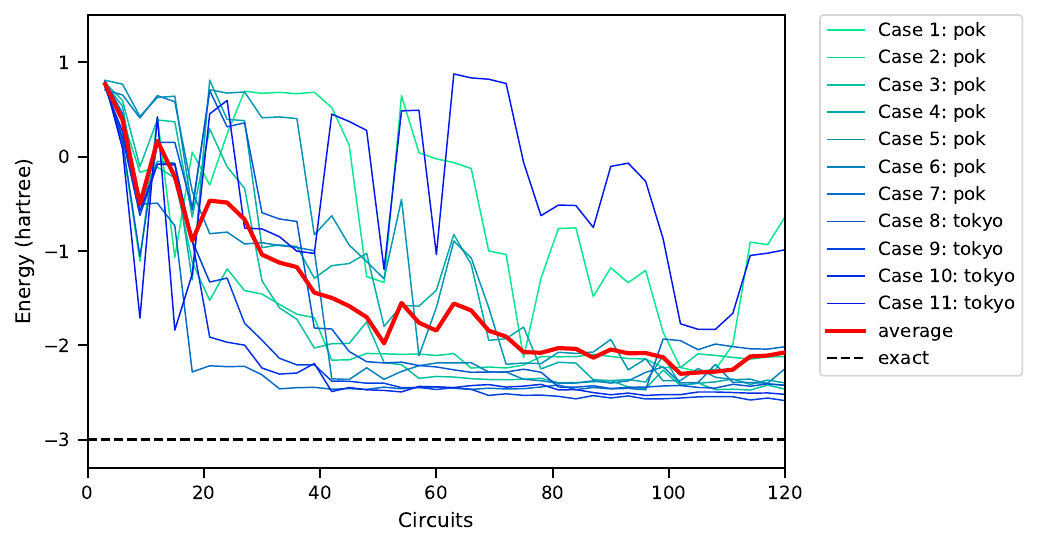}
    \caption{No-grouping without error mitigation.}
  \end{subfigure}
  \caption{%
    Experimental results of VQE on quantum computers.
    In the labels, ``pok'' and ``tokyo'' denote that the experiments were conducted on the IBM Q Poughkeepsie and IBM Q Tokyo, respectively.
    The optimizer SPSA outputs plus and minus values, but we plotted only the mean value of plus and minus values.
    The label ``average'' is the average of all cases.
    These experiments were conducted from June 4 to July 14, 2019.
}\label{fig:all_exp}
\end{figure}

We executed VQE on quantum computers several times (including the result of VQE experiment with entangled measurements of the main text) and plotted the results as shown in Fig.~\ref{fig:all_exp}.
In the experiments on VQE,
it took from several minutes to several hours to run a circuit on the quantum computers
depending on the waiting time of the job queue.
Thus, the experiments took a long time to execute all circuits.
The results depend on the device conditions, and the condition may change during the waiting time.
Although the energy failed to converge in a few experiments,
they converged faster when using the proposed method than when using the baseline
in the cases where both methods converged.

In some cases, the energy went down once and then went up again.
Thus, the converged values are not always the minimum.
This implies that SPSA is not robust enough to the noise,
and we need to try more optimizers to determine a robust one.
We also need to save the states of the optimizer as the checkpoint
and go back to the checkpoint if such fluctuation is detected.

\end{document}